\documentclass[final,1p,times,round]{elsarticle}

\usepackage{algorithm}  
\usepackage{algorithmic}  
\usepackage{bm}
\usepackage{ bbold }
\usepackage{multirow}
\usepackage{caption}
\usepackage{subfigure}
\usepackage{booktabs}
\usepackage{float}
\usepackage{booktabs}

\usepackage{color}
\usepackage{amsmath}
\usepackage{ dsfont }
\usepackage{ upgreek }
\usepackage{booktabs}
\usepackage{multirow}
\usepackage{xcolor}
\usepackage{colortbl}
\usepackage{subfigure}
\usepackage{threeparttable}



\begin{document}

\title{One-off Negative Sequential Pattern Mining}


\author[fristaddress]{Youxi Wu}
		
\ead{wuc@scse.hebut.edu.cn}

\author[fristaddress]{Mingjie Chen}
			
\author[secondaddress]{Yan Li}		

\author[fristaddress]{Jing Liu \corref{mycorrespondingauthor}}
\cortext[mycorrespondingauthor]{Corresponding author: Jing Liu}
\ead{liujing@scse.hebut.edu.cn}
			
\author[thridaddress]{Zhao Li}

\author[fifthaddress]{Jinyan Li}
			
\author[sixthaddress]{Xindong Wu}
\ead{xwu@hfut.edu.cn}
%

\author{Jing Liu}

\author{Zhao Li}

\author{Jinyan Li}

\author{Xindong Wu}

\address[fristaddress]{School of Artificial Intelligence, Hebei University of Technology, Tianjin 300401, China}
\address[secondaddress]{School of Economics and Management, Hebei University of Technology, Tianjin 300401, China}
\address[thirdaddrtess]{Alibaba-ZJU Joint Research Institute of Frontier Technologies,  Zhejiang University, Hangzhou 310000, China}
\address[fifthaddress]{Data Science Institute, University of Technology Sydney, Australia}
\address[sixthaddress]{Key Laboratory of Knowledge Engineering with Big Data (the Ministry of Education of China), Hefei University of Technology, Hefei 230009, China}




\begin{abstract}
	
Negative sequential pattern mining (SPM) is an important SPM research topic. Unlike positive SPM, negative SPM can discover events that should have occurred but have not occurred, and it can be used for financial risk management and fraud detection. However, existing methods generally ignore the repetitions of the pattern and do not consider gap constraints, which can lead to mining results containing a large number of patterns that users are not interested in. To solve this problem, this paper discovers frequent one-off negative sequential patterns (ONPs). This problem has the following two characteristics. First, the support is calculated under the one-off condition, which means that any character in the sequence can only be used once at most. Second, the gap constraint can be given by the user. To efficiently mine patterns, this paper proposes the ONP-Miner algorithm, which employs  depth-first and backtracking strategies to calculate the support. Therefore, ONP-Miner can effectively avoid creating redundant nodes and parent-child relationships. Moreover, to effectively reduce the number of candidate patterns, ONP-Miner uses pattern join and pruning strategies to generate and further prune the candidate patterns, respectively. Experimental results show that ONP-Miner not only improves the mining efficiency, but also has better mining performance than the state-of-the-art algorithms. More importantly, ONP mining can find more interesting patterns in traffic volume data to predict future traffic.  
	
\end{abstract}

%

%

\maketitle

\section{Introduction}
%
%
%
%

Sequential pattern mining (SPM) \cite{gan 1survey2019,Phi 2survey2017}, as an important knowledge discovery methods\cite{wuxindong2022,pengfeizhang,wanglizhen2022}, focuses on finding interesting subsequences (called patterns) in sequences or sequence databases (SDBs). Various SPM methods have been investigated, such as SPM with gap constraints \cite{Wu 2nonNOSEP2018,Wu 1nonHANP2021}, high-utility SPM \cite{Tin 2high2021,weisong2021, wenshenggan2021},  contrast SPM \cite{Wu 4con2021,Qing 2con2020,Rong 1con2019,David 3con2017}, order-preserving SPM for time series \cite{Hu OPP2022}, and closed SPM \cite{Jerry 1close2021,Bac 2close2017,Wu 3close2020}. However, most of these studies focuses on mining events (with each event corresponding to a character) that have occurred, which is called positive SPM \cite{Zeng 1pos2019,Ting 2pos2020}. Unlike positive SPM, negative SPM \cite{Cao 1neg2021,Cao 2neg2019} focuses on mining the events that should have occurred but have not occurred. Negative SPM has important applications in many fields, such as behavior analysis, medical services, financial risk management, and fraud detection. For example, researchers found an interesting pattern in campus data \cite{Xu campus2018}: this pattern can be written as \textbf{p} = ¬ab¬c¬dY, where `a' is Mon-breakfast, `b' is Tue-breakfast, `c' is Wed-breakfast, `d' is Thu-breakfast, and `Y' is poor grades. Thus, \textbf{p} represents students do not eat breakfast on Mondays, Wednesdays, and Thursdays, but do eat breakfast on Tuesdays, and have poor grades. Therefore, it is concluded that students who eat breakfast irregularly have poor grades. 

The current researches on negative SPM mainly focus on  multiple supports \cite{Xu msNSP2017}, progressive pattern mining \cite{Jen progre2020}, etc. However, there are two problems with these negative SPM methods. First, in practical applications, the existing negative SPM studies only consider whether a pattern occurs in a sequence, and pay less attention to the number of occurrences of a pattern in a sequence, i.e., the repetition of the patterns \cite{Wu 2nonNOSEP2018, pmdb2022}. For example, if a pattern occurs 10 times in a certain sequence, but a mining method concludes that it occurs once, then this mining method will miss a lot of important information. Second, the gap constraint is not considered, which will lead to the discovery of many meaningless patterns. An illustrative example is as follows.

\textbf{Example 1.}\label{exmp1}
Suppose we have a sequence 
$\textbf{s} $ = $\textit{s}_{1}$$ \textit{s}_{2}$$ \textit{s}_{3} $$ \textit{s}_{4}$$\textit{s}_{5}$$\textit{s}_{6} $$ \textit{s}_{7}$$\textit{s}_{8}$$ = $AACACCTC, a pattern \textbf{p} = $\textit{p}_{1}$[0,1]$\textit{p}_{2}$[0,1]¬$\textit{e}_{1}\textit{p}_{3}$ = A[0,1]C[0,1]¬GC.

Firstly, we briefly explain gap constraints [0,1] and negative character ¬G in pattern \textbf{p}. Subpattern A[0,1]C of pattern \textbf{p} means that there can be 0 or 1 character between `A' and `C'. For example, subsequence `AAC' in \textbf{s} can be matched with subpattern A[0,1]C. Moreover, as a negative character ¬G in subpattern C[0,1]¬GC  means that there can be any 0 or 1 character between `C' and `C', except `G', i.e. `G' cannot occur between `C' and `C'. For example, subsequence `CAC' can be matched with subpattern C[0,1]¬GC, while subsequence `CGC' cannot.

We know that subsequence `AACAC' in \textbf{s} can be matched with pattern \textbf{p} = A[0,1]C[0,1]¬GC. Thus, pattern \textbf{p} occurs once in sequence \textbf{s}, since classical negative SPM methods consider only whether a pattern matches with a sequence and do not consider the number of occurrences, i.e., if one or more subsequences can be matched with a pattern, pattern \textbf{p} is considered to occur once in sequence \textbf{s}, otherwise it occurs zero times. Hence, the support of pattern \textbf{p} in sequence \textbf{s} is 1 according to classical negative SPM methods.

However, if the repetition of the pattern and gap constraints are considered, the number of occurrences of a pattern will be the number of subsequences that meet the gap constraint. For example, we know that pattern \textbf{p} = A[0,1]C[0,1]¬GC matches subsequence `AACAC' in \textbf{s}. Similarly, pattern \textbf{p} also matches subsequence `ACCTC' in \textbf{s}. Hence, pattern \textbf{p} occurs twice in sequence \textbf{s} according to our method, i.e. the support of pattern \textbf{p} in sequence \textbf{s} is 2.

It can be seen from Example \ref{exmp1} that compared with classical negative SPM, the new method can find the specific occurrences of a pattern according to a gap constraint. Therefore, more valuable patterns can be discovered. In view of this, this paper proposes the problem of mining one-off negative sequential patterns (ONPs). The contributions  are as follows. 

\begin{enumerate}
	\item To avoid ignoring the repetition of a pattern, as in classical negative SPM, this paper addresses ONP mining and proposes the ONP-Miner algorithm, which involves two key steps, support calculation and pattern generation.
	
	\item To improve its efficiency, ONP-Miner adopts the depth-first and backtracking strategies to calculate the support and can effectively avoid creating useless nodes and parent-child relationships.
	
	\item To effectively reduce the number of candidate patterns, ONP-Miner employs pattern join and pruning strategies to generate and further prune the candidate patterns, respectively. 
	
	\item The experimental results show that the ONP-Miner algorithm not only has better mining performance but can also mine more valuable patterns than other competitive algorithms. More importantly, ONP mining can find more interesting patterns in traffic volume data to predict future traffic.
	
\end{enumerate}

The structure of this paper is as follows. Section 2 summarizes the related work. Section 3 gives the definition of the problem. Section 4 introduces the steps of the algorithm and shows some illustrative examples. Section 5 validates the performance of the ONP-Miner algorithm. Section 6 presents the conclusions.

\section{Related work}

Classical SPM focuses on mining frequent patterns, which means that the supports of these patterns are no less than the minimum threshold. To compress frequent patterns, Top-k SPM \cite{Chunkai 4top-k2021,Jen 5top-k2019}, closed SPM \cite{Fabio 4close2016}, and maximal SPM \cite{Md 1max2018,Yan 2max2021} were investigated. With the development of SPM, various SPM methods have been investigated for different mining tasks, such as three-way SPM \cite{Fan 1tri2020,Wu 2tri2021}, weak-gap SPM \cite{wu2022tmis,wu2022ins}, high utility SPM \cite{Yoo 3high2021,Yoo 4high2021}, spatial co-location pattern mining \citep{Wang2018colocation}, order-preserving SPM \cite{Hu OPP2022},  and contrast SPM \cite{Wu 4con2021, Qing 2con2020}. For example, three-way SPM can effectively improve the mining speed and avoid large deviations by dividing the characters into three types: strong, medium, and weak. High-utility SPM considers the external utility of the pattern, and it has been applied to various databases, such as  e-commerce databases \cite{wu2021eswa} and dynamic databases \cite{Unil dyn2020}. 

However, classical SPM methods ignore the repetition of patterns, which will cause the missing of important information. To avoid ignoring the repetition of patterns, repetitive SPM methods were studied. Repetitive SPM can be divided into mining consecutive subsequences \cite{Jin assoc2020} and mining disconsecutive subsequences \cite{Sum  inter2014}. In a consecutive subsequence, each character in the subsequence must to be continuous. Obviously, this method will miss much important information, since it is too strict. Compared with the consecutive subsequence, the disconsecutive subsequence is more flexible. 

However, if there is no constraint, then it will discover many meaningless patterns. Thus, gap constraint SPM methods were proposed. Gap constraint SPM methods not only are difficult to solve, but also have many forms, such as SPM with periodic wildcard gaps \cite{Wu 1rep2014}, nonoverlapping SPM \cite{Wu 2nonNOSEP2018,Wamg 3non2022},  and one-off SPM \cite{Fei  gap2014}. Gap constraint SPM methods have many applications, such as feature extraction for sequence classification \cite{Wu 4con2021}, keyphrase extraction  \cite{Fei  gap2014}, and finding possible promoter binding sites in DNA sequences \cite{TCBB2020gap}.  

The above studies belong to the positive SPM category, since they focus on mining the subsequences (with each character in a subsequence corresponding to an event) that have occurred. However, many scholars have discovered that missing data in a sequence may also contain important information \cite {ganneg2022}. Therefore, negative SPM has also emerged, and many negative SPM algorithms have been proposed \cite{Dong  1neg2019,Gao  2neg2021,Ping  3neg2021}. For example,  Cao et al. \cite{Cao  6neg2016ensp} proposed e-NSP, which can mine negative patterns without rescanning the database to improve efficiency. Xu et al. \cite{Xu campus2018} proposed HUNSPM, which can mine high utilities negative patterns. Huang et al. \cite{Jen progre2020} proposed Propone, which can effectively mine frequent negative patterns in progressive databases to discover the most up-to-date information.

However, the above methods ignore the repetition of the patterns. Thus, they fail to record the occurrences of the patterns. Therefore, some important information is ignored. Although Dong et al. \cite{Dong  7neg2020ernsp} focused on repetitive negative pattern mining, they did not consider gap constraints, which makes it difficult to mine negative patterns in long sequences. 

To the best of our knowledge, e-NSP \cite{Cao  6neg2016ensp} and e-RNSP \cite{Dong  7neg2020ernsp} are the closest negative SPM methods. Since they do not consider the repetition of patterns and gap constraints, the two methods generate negative candidate patterns by converting some positive characters. To overcome the drawbacks of these methods, this paper proposes the ONP-Miner algorithm to perform one-off negative SPM, which considers the repetition of patterns and gap constraints.

\section{Problem definition}

A sequence with \textit{n} events can be expressed as \textbf{s} = $\textit{s}_{1}$$ \textit{s}_{2}$...$ \textit{s}_{i}$...$\textit{s}_{n}$(1 $\leq$ \textit{i} $\leq$ \textit{n}), $\textit{s}_{i}\in\Sigma$, where \textit{n} is the length of \textbf{s}, and $\Sigma$ represents a set of characters in sequence \textbf{s}. For example,  in a DNA sequence \textbf{s} = $\textit{s}_{1}$$ \textit{s}_{2}$$ \textit{s}_{3} $$ \textit{s}_{4}$$\textit{s}_{5}$$\textit{s}_{6}$$ = $ATTACG,  the character set is $\Sigma$ = \{A, G, C, T\}, and the length is \textit{n} = 6. A pattern with gap constraints can be expressed as \textbf{p} = $\textit{p}_{1}$[\textit{M},\textit{N}]$\textit{p}_{2}$...[\textit{M},\textit{N}]$\textit{p}_{j}$...[\textit{M},\textit{N}]$\textit{p}_{m}$(1 $\leq$ \textit{j} $\leq$ \textit{m}, 0 $\leq$ \textit{M} $\leq$ \textit{N}), where \textit{m} is the length of \textbf{p}, and \textit{M} and \textit{N} refer to the minimum and maximum number of wildcards allowed between $\textit{p}_{j-1}$ and $\textit{p}_{j}$, respectively.

\textbf{Definition 1.}
	\label{def1}
If \textit{L} = <$\textit{l}_{1}$,$ \textit{l}_{2}$,...,$ \textit{l}_{j}$,...,$ \textit{l}_{m}$> is a set of integers satisfying $\textit{p}_{1}$ =$\textit{s}_{l_{1}}$, $\textit{p}_{2}$ =$\textit{s}_{l_{2}}$, ..., $\textit{p}_{m}$ =$\textit{s}_{l_{m}}$, and \textit{M} $\leq$ $ \textit{l}_{j+1}$-$ \textit{l}_{j}$-1 $\leq$ \textit{N}, then \textit{L} is an occurrence of \textbf{p} in \textbf{s}.

\textbf{Definition 2.}
	\label{def2}
Suppose \textit{L} = <$\textit{l}_{1}$,$ \textit{l}_{2}$,...,$ \textit{l}_{k}$,...,$ \textit{l}_{m}$> and  \textit{L'} = <$\textit{l'}_{1}$,$ \textit{l'}_{2}$,...,$ \textit{l'}_{j}$,...,$ \textit{l'}_{m}$> are two occurrences of \textbf{p} in sequence \textbf{s}. If $ \textit{l}_{k}$=$ \textit{l'}_{j}$(1 $\leq$ \textit{k} $\leq$ \textit{m}, 1 $\leq$ \textit{j} $\leq$ \textit{m}), \textit{L} and \textit{L'} do not satisfy the one-off condition, otherwise, \textit{L} and \textit{L'} do satisfy the one-off condition.  
	
\textbf{Example 2.}
\label{exmp2}
Suppose we have a sequence \textbf{s} = AACACCTC and a pattern \textbf{p} = A[0,1]C[0,1]C. According to the gap constraint [0,1], all occurrences of \textbf{p} in \textbf{s} are <1,3,5>,<2,3,5>,<4,5,6>, and <4,6,8>. <1,3,5> and <4,5,6> do not meet the one-off condition, since 5 is used in these two occurrences, while <1,3,5> and <4,6,8> do satisfy the one-off condition.
	
\textbf{Definition 3.}
\label{def3}
Given a character \textit{e} ($\textit{e} \in \Sigma$ or NULL), its corresponding negative character is ¬\textit{e}, which represents the missing character \textit{e}. A negative pattern with gap constraints can be expressed as \textbf{p}=$\textit{p}_{1}$[\textit{M},\textit{N}]¬$\textit{e}_{1}$$\textit{p}_{2}$...[\textit{M},\textit{N}]¬$\textit{e}_{j-1}$$\textit{p}_{j}$...[\textit{M},\textit{N}] ¬$\textit{e}_{m-1}$$\textit{p}_{m}$(1 $\leq$ \textit{j} $\leq$ \textit{m}):
\begin{enumerate}
	\item If $\textit{e}_{j-1} \in \Sigma$, then $\textit{p}_{j-1}$[\textit{M},\textit{N}]¬$\textit{e}_{j-1}$$\textit{p}_{j}$ is a negative subpattern  which means that $\textit{e}_{j-1}$ does not exist between $\textit{s}_{l_{j-1}}$ and $\textit{s}_{l_{j}}$.
	
	\item If $\textit{e}_{j-1}$ is NULL, then $\textit{p}_{j-1}$[\textit{M},\textit{N}]$\textit{p}_{j}$ is a classical positive subpattern.
\end{enumerate}

\textbf{Definition 4.}
\label{def4}
The support of pattern \textbf{p} in sequence \textbf{s} is the number of occurrences under the one-off condition, and it is denoted by \textit{sup}(\textbf{p},\textbf{s}). The support of \textbf{p} in sequence database \textit{SDB}, denoted by $\textit{sup}(\textbf{p}, \textit{SDB})$, is the sum of the support of each sequence, i.e., \textit{sup}(\textbf{p}, \textit{SDB}) = $\textstyle \sum_{k=1}^{N} \textit{sup}(\textbf{p},\textbf{s}_\textit{k})$. If the support of \textbf{p} in \textbf{s} or in \textit{SDB} is no less than the minimum support threshold \textit{minsup}, then \textbf{p} is called a frequent ONP.

\textbf{Definition 5.}
\label{def5}
Given a minimum support threshold \textit{minsup}, our objective is to mine all frequent ONPs in a sequence or sequence database.

\textbf{Example 3.}
\label{exmp3}
Suppose we have a sequence \textbf{s} = $\textit{s}_{1}$$ \textit{s}_{2}$$ \textit{s}_{3} $$ \textit{s}_{4}$$\textit{s}_{5}$$\textit{s}_{6} $$ \textit{s}_{7}$$\textit{s}_{8}$$ = $AACACCTC, \textit{minsup} = 2, \textit{gap} = [0,1], and $\Sigma$ = \{A, C, G, T\}.

<1,3,5> is an occurrence of \textbf{p} = $\textit{p}_{1}$[\textit{M},\textit{N}]$\textit{p}_{2}$[\textit{M},\textit{N}]¬$\textit{e}_{1}$$\textit{p}_{3}$ = A[0,1]C[0,1]¬GC, since $\textit{s}_{1}$ = ``A'' = $\textit{p}_{1}$, $\textit{s}_{3}$ = ``C'' = $\textit{p}_{2}$, $\textit{s}_{5}$= ``C'' = $\textit{p}_{3}$, and $\textit{s}_{4}$ = ``A''$\ne$ $\textit{e}_{1}$. Similarly, another occurrence is <4,6,8>. Thus, \textit{sup}(\textbf{p}, \textbf{s})=2$\geq$\textit{minsup}. Therefore, \textbf{p} = A[0,1]C[0,1]¬GC is a frequent ONP. Furthermore, all of the frequent ONPs follow: \{A, C, A[0,1]C, C[0,1]C, A[0,1]¬GC, A[0,1]¬TC, C[0,1]¬CC, C[0,1]¬GC, A[0,1]C[0,1]C, A[0,1]C[0,1]¬CC, A[0,1]¬GC[0,1]¬CC, A[0,1]¬TC[0,1]¬CC, A[0,1]C[0,1]¬GC, A[0,1]¬GC[0,1]¬GC, A[0,1]¬TC[0,1]¬GC, A[0,1]¬GC[0,1]C, A[0,1]¬TC[0,1]C\}. 

\section{Proposed algorithm}

There are two key issues in our research: support calculation and candidate pattern generation. Section 4.1 proposes the MatchDB algorithm to calculate the support. Section 4.2 introduces the candidate generation methods. Section 4.3 presents the ONP-Miner algorithm, and analyzes the time and space complexities.

\subsection{Nettree and MatchDB algorithm}

In this section, we first briefly review the selection of the rightmost parent (SRMP) algorithm \cite{Yu  SRMP2021}, which is a support calculation method based on the nettree structure \cite{Wu  nettree2017}. To improve its efficiency, we propose an efficient matching algorithm based on the depth-first and backtracking strategies, named MatchDB. 

To calculate the support, SRMP first creates a nettree according to sequence \textbf{s} and pattern \textbf{p}, and then it uses a rightmost parent strategy to search for an occurrence. An illustrative example follows.

\begin{figure}[]
	\centering
	\includegraphics[width=0.4\linewidth]{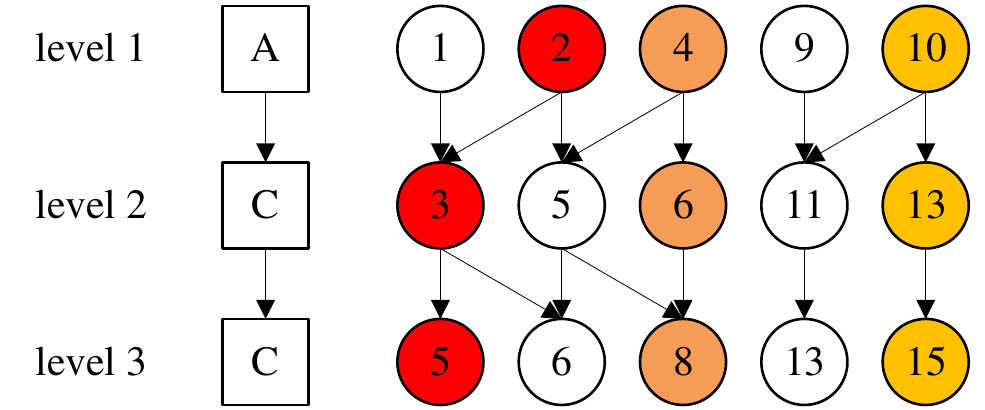}
	\caption{A nettree of \textbf{p} in \textbf{s}. Unlike a tree, a nettree may have more than one root and a node may have more than one parent. For example, the nettree has five roots, $n_{1}^{1}$, $n_{1}^{2}$, $n_{1}^{4}$, $n_{1}^{9}$, and $n_{1}^{10}$. Node $n_{2}^{5}$ has two parents, $n_{1}^{2}$ and $n_{1}^{4}$.}
	\label{fig:figure1}
\end{figure}

\textbf{Example 4.}
\label{exmp4}
Suppose we have a sequence \textbf{s} = $\textit{s}_{1}$$ \textit{s}_{2}$$ \textit{s}_{3} $$ \textit{s}_{4}$$\textit{s}_{5}$$\textit{s}_{6} $$ \textit{s}_{7}$$\textit{s}_{8}$$\textit{s}_{9}$$\textit{s}_{10}$$\textit{s}_{11}$$\textit{s}_{12}$$\textit{s}_{13}$$\textit{s}_{14}$$\textit{s}_{15}$ = AACACCTCAACGCTC and a pattern \textbf{p} = $\textit{p}_{1}$[\textit{M},\textit{N}]$\textit{p}_{2}$[\textit{M},\textit{N}]$\textit{p}_{3}$=A[0,2]C[0, 2]C. SRMP creates a nettree as shown in Fig. ~\ref{fig:figure1}. We know that the largest leaf is $n_{3}^{15}$. Thus, we get a one-off occurrence <10,13,15>, since the rightmost parent node of $n_{3}^{15}$ is $n_{2}^{13}$, whose rightmost parent node is $n_{1}^{10}$. Similarly, SRMP obtains other two one-off occurrences, <4,6,8> and <2,3,5>.

\begin{figure}[]
	\centering
	\includegraphics[width=0.6\linewidth]{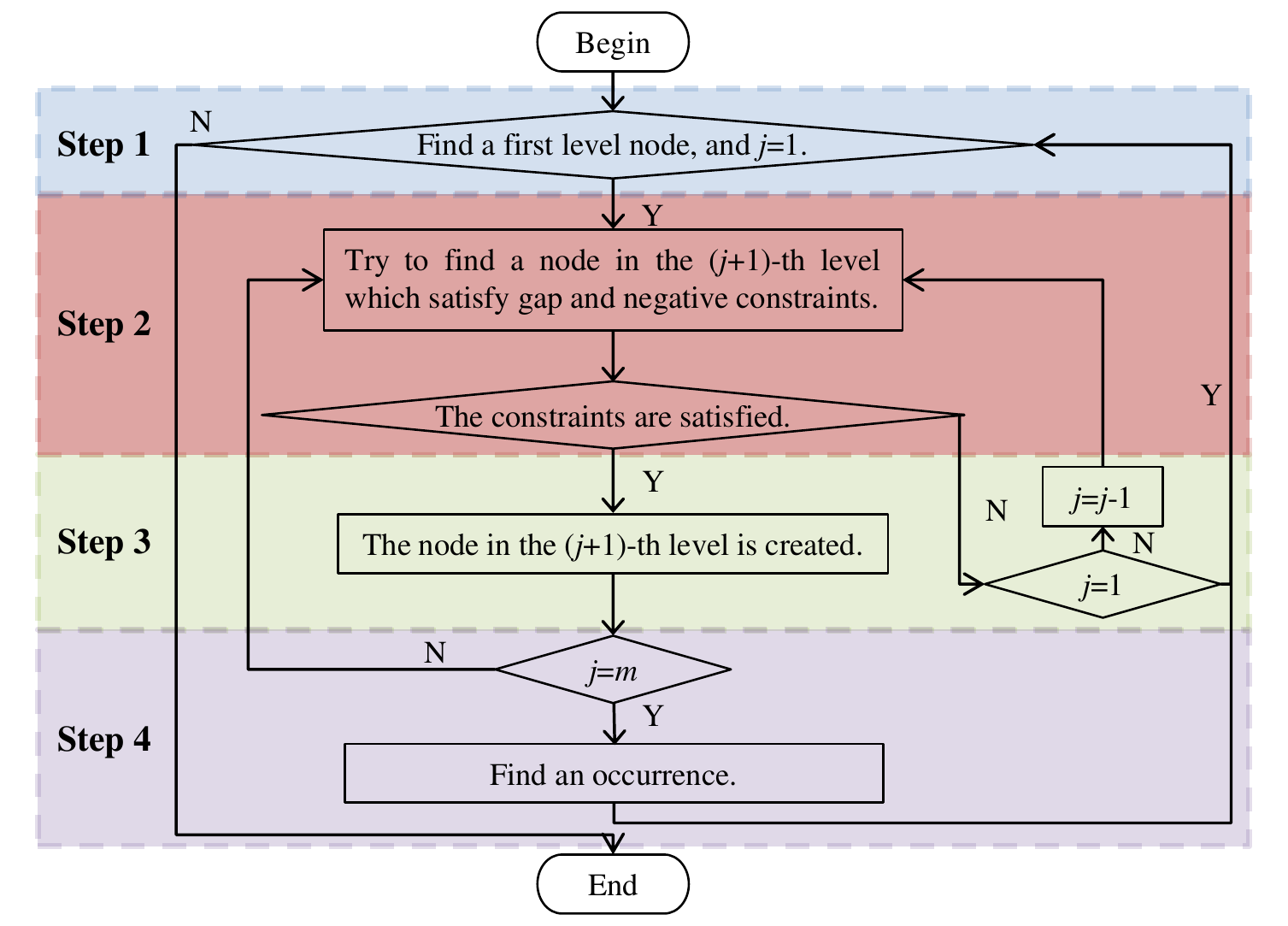}
	\caption{The  flowchart of MatchDB.}
	\label{fig:figure2}
\end{figure}

Although SRMP can calculate the support, it has two issues. First, SRMP does not consider the support calculation of negative patterns. More importantly, SRMP is not efficient, since it creates many useless nodes and parent-child relationships. To tackle these two issues, we propose an efficient MatchDB algorithm, which employs the depth-first and backtracking strategies to calculate the support of a negative sequence. The flowchart of MatchDB is shown in Fig. ~\ref{fig:figure2}.

\textbf{Step 1}: If $s_{i}$ = $p_{1}$, then node $n_{1}^{i}$ is created as a root in the first level.

\textbf{Step 2}: Suppose that node $n_{j}^{t}$ has been created in the $j^{\mathrm{th}}$ level. Now, MatchDB will find the child node of node $n_{j}^{t}$. Suppose that $s_{i_{j+1}}$ = $p_{j+1}$ (\textit{j} $\geq$ 1), \textit{t} and $i_{j+1}$ satisfy the gap constraint [\textit{M},\textit{N}], i.e., \textit{M}$\leq$$i_{j+1}$-\textit{t}-1$\leq$\textit{N}, $s_{i_{j+1}}$ is not used by another occurrence, and $n_{j+1}^{i_{j+1}}$ has not been created. There are two cases.

\textbf{Case 1}: If ¬$e_{j}$ is NULL, then node$n_{j+1}^{i_{j+1}}$ is created in the $(j+1)^{\mathrm{th}}$ level, and a parent-child relationship is created between nodes $n_{j}^{t}$ and $n_{j+1}^{i_{j+1}}$.

\textbf{Case 2}: If ¬$e_{j}$ is not NULL and all characters between $s_{t}$ and $s_{i_{j+1}}$ are different from $e_{j}$, then node $n_{j+1}^{i_{j+1}}$ is created in the $(j+1)^{\mathrm{th}}$ level, and a parent-child relationship is created between nodes $n_{j}^{t}$ and $n_{j+1}^{i_{j+1}}$.

\textbf{Step 3}: If a node is created in Step 2, then MatchDB iterates Step 2, until node $n_{m}^{i_{m}}$ in the $m^{\mathrm{th}}$ level is created. If no node is created and \textit{j}=1, then MatchDB iterates Step 1 to find a new root. If there is no node to be created and \textit{j}>1, then MatchDB iterates Step 2 to create its child.

\textbf{Step 4}: If node $n_{m}^{i_{m}}$ is created, then MatchDB find an occurrence $<$$s_{i_{1}}$, $s_{i_{2}}$,…,  $s_{i_{m}}$$>$, which cannot be reused. It iterates Step 1 to find a new occurrence, until no new root is generated. 

Example 5 illustrates the principle of MatchDB.

\begin{figure}[]
	\centering
	\includegraphics[width=0.4\linewidth]{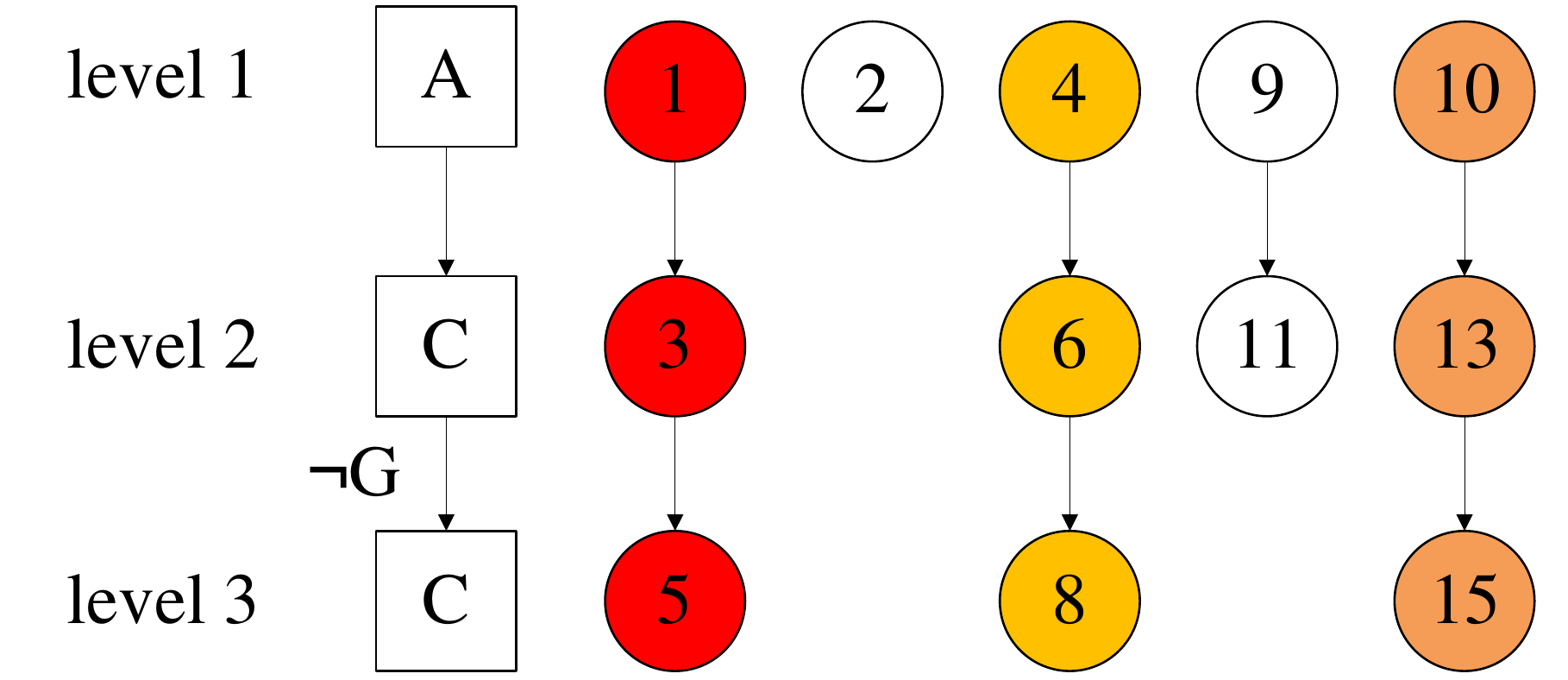}
	\caption{Matching process for \textbf{p} in \textbf{s} according to the MatchDB algorithm.}
	\label{fig:figure3}
\end{figure}

\textbf{Example 5.} \label{exmp5}
Suppose we have a sequence \textbf{s} = $\textit{s}_{1}$$ \textit{s}_{2}$$ \textit{s}_{3} $$ \textit{s}_{4}$$\textit{s}_{5}$$\textit{s}_{6} $$ \textit{s}_{7}$$\textit{s}_{8}$$\textit{s}_{9}$$\textit{s}_{10}$$\textit{s}_{11}$$\textit{s}_{12}$$\textit{s}_{13}$$\textit{s}_{14}$$\textit{s}_{15}$ = AACACCTCAACGCTC and a pattern \textbf{p} = A[0,2]C[0,2]¬GC. The matching process is shown in Fig. ~\ref{fig:figure3}. 

Suppose that MatchDB obtains the first occurrence <1,3,5>. Now, MatchDB finds the next occurrence. It creates a root $n_{1}^{2}$, since $\textit{s}_{2}$ = A = $\textit{p}_{1}$. It creates a child node of $n_{1}^{2}$ according to ``A[0,2]C''. Although $\textit{s}_{3}$ = $\textit{s}_{5}$ = C = $\textit{p}_{2}$, $\textit{s}_{3}$ and $\textit{s}_{5}$ have been used in occurrence <1,3,5>. Therefore, $n_{1}^{2}$ does not have a child. MatchDB backtracks to the parent node of $n_{1}^{2}$. Thus, MatchDB finds a new root $n_{1}^{4}$. It is clear that MatchDB obtains an occurrence <4,6,8>.

Now, MatchDB finds the next root $n_{1}^{9}$. MatchDB iterates Step 2 and finds the child node of $n_{1}^{9}$ which is $n_{2}^{11}$. Although $\textit{s}_{13}$ = $\textit{p}_{3}$, node $n_{3}^{13}$ cannot be created, since $\textit{s}_{12}$ = G which is between nodes $\textit{s}_{11}$ and $\textit{s}_{13}$. MatchDB backtracks to node $n_{1}^{9}$ to find another child. Since node $n_{1}^{9}$ does not have another child, MatchDB finds a new root $n_{1}^{10}$. Although $\textit{s}_{11}$ = C = $\textit{p}_{2}$ and 10 and 11 satisfy the gap constraint [0,2], $n_{2}^{11}$ cannot be selected, since $n_{2}^{11}$ is a child of $n_{1}^{9}$. MatchDB creates $n_{2}^{13}$ as a child of node $n_{1}^{10}$. Since $\textit{s}_{14}$ = T $\neq$ G and $\textit{s}_{15}$ = C = $\textit{p}_{3}$, C[0,2]¬GC is satisfied and $n_{3}^{15}$ is created. Thus, MatchDB obtains the third occurrence <10,13,15>.

From Example 5, it can be seen that MatchDB employs the depth-first and backtracking strategies to find the occurrences, as shown in Algorithm 1.

\begin{algorithm}[]
	\label{Algom 1}
	\caption{MatchDB} 
	\hspace*{0.02in} \leftline{{\bf Input:} Sequence database \textit{SDB} and pattern \textbf{p}}\\ 
	\hspace*{0.02in} \leftline{{\bf Output:} \textit{sup}(\textbf{p}, \textit{SDB})}\\
	\begin{algorithmic}[1]
			\STATE \textit{sup}(\textbf{p}, \textit{SDB})←0; 
			\FOR{each \textbf{s} in \textit{SDB}}	
			\STATE \textit{n}←\textbf{s}.length();
		    \FOR{\textit{i}=1 to \textit{n}-\textit{m}+1}	
		    \IF{$\textit{s}_{i}$=$\textit{p}_{1}$ and \textit{i} is not used}
		    \STATE Create node $n_{1}^{i}$, \textit{occ}[1]←$n_{1}^{i}$;
		    \STATE occ←DFB(\textbf{s}, \textbf{p}, \textit{occ}[1]);
		    \IF{\textit{occ}$\neq$NULL}
		    \STATE \textit{sup}(\textbf{p}, \textbf{s})++;
			\STATE	 All nodes in occ cannot be reused;
		    \ENDIF
		    \ENDIF
			\ENDFOR		
			\ENDFOR	
			\STATE return \textit{sup}(\textbf{p}, \textit{SDB});
	\end{algorithmic}
\end{algorithm}
MatchDB employs DFB to perform the depth-first and backtracking strategies to calculate the support, as shown in Algorithm 2.

\begin{algorithm}[]
	\label{Algom 2}
	\caption{DFB} 
	\hspace*{0.02in} \leftline{{\bf Input:} Sequence \textbf{s}, pattern \textbf{p}, \textit{occ}[1]}\\
	\hspace*{0.02in} \leftline{{\bf Output:} \textit{occ}}\\
	\begin{algorithmic}[1]
		\STATE \textit{l}←1
		\WHILE{\textit{l}<\textit{m} }
		\STATE \textit{child}←find the first child of \textit{occ}[\textit{l}];
		\IF{\textit{child}=NULL}
		\STATE \textit{l}←\textit{l}-1;
		\IF{\textit{l}=0}
		\STATE return NULL;
		\ENDIF
		\ELSE
		\STATE \textit{occ}[\textit{l}+1]←\textit{child}, \textit{l}←\textit{l}+1;
		\ENDIF
		\ENDWHILE
		\STATE return \textit{occ};
	\end{algorithmic}
\end{algorithm}

\subsection{Candidate pattern generation}

To effectively generate candidate patterns, we propose a method that has two strategies: pruning and pattern join. We first prove that the support of a negative sequential pattern is not greater than the support of its corresponding positive sequential pattern. Then, we propose a pruning strategy to reduce the candidate patterns. Finally, we adopt a pattern join strategy to generate candidate patterns. 

We know that the existing negative SPM methods do not consider the repetition of a pattern in a sequence which means that if a pattern occurs in a sequence, then the support of the pattern in the sequence database will plus one. Therefore, the existing negative SPM proved that it meets the anti-monotonicity in a database. However, our problem is different from the existing negative SPM, since our problem considers the repetition of a pattern in a sequence. As a new problem, we prove that our problem also meets the anti-monotonicity in a sequence.

\textbf{Theorem 1.}
\label{Theorem1}
The support of a one-off negative sequential pattern is not greater than the support of its corresponding one-off positive sequential pattern.

\textbf{Proof}: Suppose we have a sequence \textbf{s} and a negative sequential pattern \textbf{p} = $\textit{p}_{1}$[\textit{M},\textit{N}]$\textit{p}_{2}$...$\textit{p}_{j-1}$[\textit{M},\textit{N}]¬$\textit{e}_{j-1}$$\textit{p}_{j}$...[\textit{M},\textit{N}]$\textit{p}_{m}$, where $\textit{e}_{j-1} \in \Sigma$. The corresponding positive sequential pattern of \textbf{p} is \textbf{q} = $\textit{p}_{1}$[\textit{M},\textit{N}]$\textit{p}_{2}$...$\textit{p}_{j-1}$[\textit{M},\textit{N}]$\textit{p}_{j}$...[\textit{M},\textit{N}]$\textit{p}_{m}$. Suppose that \textit{L} = <$\textit{l}_{1}$,$ \textit{l}_{2}$,...,$ \textit{l}_{j-1}$,$ \textit{l}_{j}$,...,$ \textit{l}_{m}$>is an occurrence of pattern \textbf{p} in \textbf{s}, which means that there is no character $\textit{e}_{j-1}$ between $\textit{s}_{l_{j-1}}$ and $\textit{s}_{l_{j}}$. For a positive sequential pattern, we do not need to consider the negative character ¬$\textit{e}_{j-1}$. Therefore, \textit{L} is also an occurrence of pattern \textbf{q} in \textbf{s}. Hence, \textit{sup}(\textbf{q}, \textbf{s}) $\geq$ \textit{sup}(\textbf{p}, \textbf{s}).

Based on Theorem 1, we know that if \textit{sup}(\textbf{q}, \textbf{s})<\textit{minsup}, then \textit{sup}(\textbf{p}, \textbf{s})<\textit{minsup}, since \textit{sup}(\textbf{q}, \textbf{s})$\geq$\textit{sup}(\textbf{p}, \textbf{s}). Thus, we propose the pruning strategy.

\textbf{The pruning strategy}: 
\label{prun}
If the corresponding positive sequence of a candidate pattern is not frequent, then the candidate pattern is not frequent and can be pruned. 

Now, we will introduce the pattern join strategy.

\textbf{Definition 6.}
\label{def6}
Suppose we have a pattern \textbf{p} = $\textit{p}_{1}$[\textit{M},\textit{N}]¬$\textit{e}_{1}$$\textit{p}_{2}$...[\textit{M},\textit{N}]¬$\textit{e}_{m-2}$$\textit{p}_{m-1}$[\textit{M},\textit{N}]¬$\textit{e}_{m-1}$$\textit{p}_{m}$, the prefix of \textbf{p} is $\textit{p}_{1}$[\textit{M},\textit{N}]¬$\textit{e}_{1}$$\textit{p}_{2}$...[\textit{M},\textit{N}]¬$\textit{e}_{m-2}$$\textit{p}_{m-1}$, denoted by \textit{Prefix}(\textbf{p}), and the suffix of \textbf{p} is $\textit{p}_{2}$...[\textit{M},\textit{N}]¬$\textit{e}_{m-2}$$\textit{p}_{m-1}$[\textit{M},\textit{N}]¬$\textit{e}_{m-1}$$\textit{p}_{m}$, denoted by \textit{Suffix}(\textbf{p}).

\textbf{Definition 7.} \label{def7}
Suppose we have two patterns \textbf{p} = $\textit{p}_{1}$[\textit{M},\textit{N}]¬$\textit{e}_{1}$$\textit{p}_{2}$...[\textit{M},\textit{N}]¬$\textit{e}_{m-2}$$\textit{p}_{m-1}$[\textit{M},\textit{N}]¬$\textit{e}_{m-1}$$\textit{p}_{m}$ and \textbf{q} = $\textit{q}_{1}$[\textit{M},\textit{N}]¬$\textit{e'}_{1}$ $\textit{q}_{2}$...[\textit{M},\textit{N}]¬$\textit{e'}_{m-2}$$\textit{q}_{m-1}$[\textit{M},\textit{N}]¬$\textit{e'}_{m-1}$$\textit{q}_{m}$. If pattern \textbf{r} = \textit{Suffix}(\textbf{p}) = \textit{Prefix}(\textbf{q}), then superpattern \textbf{t} = \textbf{p}$\oplus$\textbf{q} = $\textit{p}_{1}$[\textit{M},\textit{N}] ¬$\textit{e}_{1}$\textbf{r}¬$\textit{e'}_{m-1}$$\textit{q}_{m}$  is generated . This method is called the pattern join strategy.

Example 6 illustrates the principle of the pattern join strategy.

\begin{figure}[]
	\centering
	\includegraphics[width=0.3\linewidth]{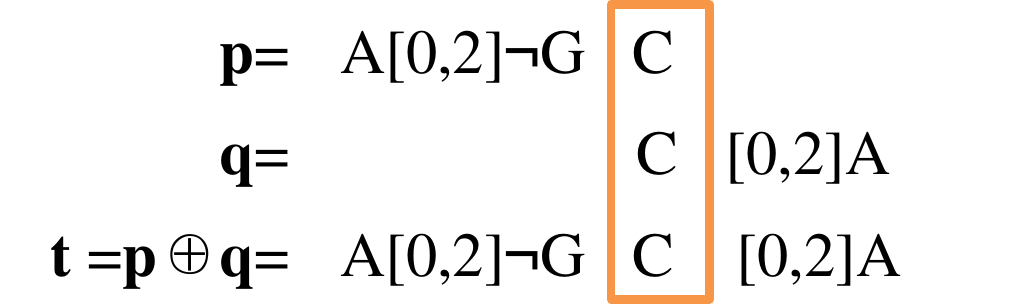}
	\caption{An illustrative example of pattern join}
	\label{fig:figure4}
\end{figure}

\textbf{Example 6.}
\label{exmp6}
Suppose we have two patterns \textbf{p} = A[0,2]¬GC and \textbf{q} = C[0,2]A. Since \textbf{r} = \textit{Suffix}(\textbf{p}) = \textit{Prefix}(\textbf{q}) = C, \textbf{p} and \textbf{q} can generate a candidate \textbf{t} = \textbf{p}$\oplus$\textbf{q} = A[0,2]¬GC[0,2]A, as shown in Fig. ~\ref{fig:figure4}. 

Now, we will use Example 7 to illustrate that the pattern join and pruning strategies outperform the enumeration tree strategy and pattern join strategy.

\textbf{Example 7.} \label{exmp7}
Suppose we have a sequence \textbf{s} = $\textit{s}_{1}$$ \textit{s}_{2}$$ \textit{s}_{3} $$ \textit{s}_{4}$$\textit{s}_{5}$$\textit{s}_{6} $$ \textit{s}_{7}$$\textit{s}_{8}$$\textit{s}_{9}$$\textit{s}_{10}$$\textit{s}_{11}$$\textit{s}_{12}$$\textit{s}_{13}$$\textit{s}_{14}$$\textit{s}_{15}$ = AACACCTCAACGCTC, \textit{gap} = [\textit{M}, \textit{N}] = [0,2], \textit{minsup} = 3, and $\Sigma$ = \{A, C, G, T\}.

The enumeration tree strategy enumerates all possible candidate patterns in a tree. Generally, we adopt the depth-first search and breadth-first search to realize the enumeration tree strategy. Suppose we have a frequent pattern \textbf{p} with length \textit{m}. If we use the enumeration tree strategy to generate the candidate patterns, there are 20 candidate patterns with length \textit{m}+1 when $\Sigma$ = \{A, C, G, T\}. The reason for this is as follows. First, there are four candidate patterns with length \textit{m}+1: \textbf{p}[\textit{M},\textit{N}]A, \textbf{p}[\textit{M},\textit{N}]C, \textbf{p}[\textit{M},\textit{N}]G, and \textbf{p}[\textit{M},\textit{N}]T. Moreover, for each candidate pattern, there are four cases in which negative characters can be added. We take \textbf{p}[\textit{M},\textit{N}]A as an example. There are four negative candidate patterns: \textbf{p}[\textit{M},\textit{N}]¬AA, \textbf{p}[\textit{M},\textit{N}]¬CA, \textbf{p}[\textit{M},\textit{N}]¬GA, and \textbf{p}[\textit{M},\textit{N}]¬TA. Hence, each frequent pattern generates 20 candidate patterns. In this example, we know that there are seven frequent patterns with length two: A[0,2]C, A[0,2]¬GC, A[0,2]¬TC, C[0,2]A, C[0,2]C, C[0,2]¬CC, and C[0,2]¬GC. Therefore, there are 140 candidate patterns with length three according to the enumeration tree strategy. 

For the pattern join strategy, there are only 27 candidate patterns with length three. For example, A[0,2]C, A[0,2]¬GC, and A[0,2]¬TC can join with C[0,2]A. Therefore, we get the following candidate patterns with length three: A[0,2]C[0,2]A, A[0,2]¬GC[0,2]A, and A[0,2]¬TC[0,2]A. 

For the pattern join and pruning strategies, we know that A[0,2]C[0,2]A is infrequent. Thus, according to Theorem 1, we know that A[0,2]¬GC[0,2]A and A[0,2]¬TC[0,2]A are also infrequent. Therefore, the pruning strategy can further prune the 27 candidate patterns generated by the pattern join strategy. There are 15 candidate patterns with length three. Table  \ref{tab1} shows a comparison of the number of candidate patterns.

\begin{table}[]
		\footnotesize
	\caption{Comparison of the number of candidate patterns\\}
	\centering
	\label{tab1}
	\begin{tabular}{@{}lllll@{}}
		\toprule
		& Length=2 & Length=3 & Length=4 & Total \\ \midrule
		Depth-first search for enumeration tree   & 80       & 140      & 180      & 400   \\
		Breadth-first search for enumeration tree & 80       & 140      & 180      & 400   \\
		Pattern join strategy                     & 20       & 27       & 1        & 48    \\
		Pattern join and pruning strategies       & 16       & 15       & 1        & 32    \\
		Frequent patterns                         & 7        & 9        & 0        & 16    \\ \bottomrule
	\end{tabular}
\end{table}

According to Table  \ref{tab1}, we know that the pattern join and pruning strategies only check 32 candidate patterns, while the pattern join, depth-first search, and breadth-first search methods check 48, 400, and 400 candidate patterns, respectively. We know that the smaller the number of candidate patterns, the faster the algorithm. Hence, the pattern join and pruning strategies outperform other strategies.

\subsection{ONP-Miner algorithm}

Based on the above content, this paper proposes the ONP-Miner algorithm for mining frequent ONPs. The main steps are as follows.

\textbf{Step 1}: Traverse the sequence database to find frequent patterns with length one.

\textbf{Step 2}: Generate the positive candidate patterns with length two. Prune the infrequent candidate patterns. Finally, generate the negative candidate patterns based on the frequent patterns and calculate the support for each pattern to find the frequent patterns with length two.

\textbf{Step 3}: Generate all candidate patterns with length \textit{l}+1 using pattern join.

\textbf{Step 4}: Calculate the support of each positive candidate pattern. If its support is no less than the threshold, store the pattern in $\textit{F}_{l+1}$; otherwise, prune its corresponding negative candidate patterns.

\textbf{Step 5}: Calculate the support of each negative candidate pattern. If its support is no less than the threshold, store the pattern in $\textit{F}_{l+1}$.

\textbf{Step 6}: Iterate Steps 3 to 5 until the set of frequent patterns is empty.

Example 8 illustrates the principle of the ONP-Miner algorithm.

\textbf{Example 8.}
\label{exmp8}
We use the same sequence and constraints in Example 7. 

Step 1: Traverse the sequence to calculate the number of occurrences for each character. Although $\Sigma$ = \{A, C, G, T\}, the frequent pattern set $\textit{F}_{1}$ = \{A, C\}, since their supports are no less than \textit{minsup} = 3.

Step 2: Generate the positive candidate patterns with length two. We know that A[0,2]A, A[0,2]C, C[0,2]A, and C[0,2]C are candidate patterns. Since the supports of A[0,2]C, C[0,2]A, and C[0,2]C are no less than \textit{minsup}, we generate the negative candidate patterns based on these three patterns. We can generate four negative candidate patterns based on each positive pattern. For example, we can generate A[0,2]¬AC, A[0,2]¬CC, A[0,2]¬GC, and A[0,2]¬TC based on A[0,2]C. Thus, there are 12 negative candidate patterns and the supports of A[0,2]¬GC, A[0,2]¬TC, C[0,2]¬CC, and C[0,2]¬GC are no less than \textit{minsup}. Hence, $\textit{F}_{2}$ = \{A[0,2]C, C[0,2]A, C[0,2]C, A[0,2]¬GC, A[0,2]¬TC, C[0,2]¬CC, C[0,2]¬GC\}.

Step 3: Generate all candidate patterns with length three using pattern join. As mentioned in Example 7, there are 27 candidate patterns, including A[0,2]C[0,2]A, A[0,2]C[0,2]C, C[0,2]A[0,2]C, A[0,2]¬GC[0,2]A, and A[0,2]¬TC[0,2]A.

Step 4: Calculate the support for each positive candidate pattern. The supports of A[0,2]C[0,2]C and C[0,2]A[0,2]C are no less than \textit{minsup}, so they are stored in $\textit{F}_{3}$. Now, $\textit{F}_{3}$ = \{A[0,2]C[0,2]C, C[0,2]A[0,2]C\}. The supports of A[0,2]C[0,2]A, C[0,2]C[0,2]A, and C[0,2]C[0,2]C are less than \textit{minsup}, and therefore their corresponding negative candidate patterns are pruned, including A[0,2]¬GC[0,2]A and A[0,2]¬TC[0,2]A.

Step 5: Calculate the support for each negative candidate pattern. The supports of A[0,2]C[0,2]¬CC, A[0,2]C[0,2]¬GC, A[0,2]¬GC[0,2]C, A[0,2]¬GC[0,2]¬CC, A[0,2]¬TC[0,2]C, A[0,2]¬TC[0,2]¬CC, and A[0,2]¬TC[0,2]¬GC are no less than \textit{minsup}, thus they are ONPs. Now, $\textit{F}_{3}$ = \{A[0,2]C[0,2]C, C[0,2]A[0,2]C, A[0,2]C[0,2]¬CC, A[0,2]C[0,2]¬GC, A[0,2]¬GC[0,2]C, A[0,2]¬GC[0,2]¬CC, A[0,2]¬TC[0,2]C, A[0,2]¬TC[0,2]¬CC, A[0,2]¬TC[0,2]¬GC\}.

Step 6: Iterate Steps 3 to 5 until the set of frequent patterns is empty; then the algorithm ends.

Thus, 16 ONPs are mined. 

The ONP-Miner algorithm is shown in Algorithm 3.
\begin{algorithm}[h]
	\label{Algom 3}
	\caption{ONP-Miner} 
	\hspace*{0.02in} \leftline{{\bf Input:} Sequence dababase \textit{SDB}, \textit{minsup}, and \textit{gap}}
	\hspace*{0.02in} \leftline{{\bf Output:} ONPs stored in \textit{F}}\\
	\begin{algorithmic}[1]
		\STATE Traverse \textit{SDB}, get $\Sigma$, and store frequent positive patterns with length one in $\textit{F}_{1}$;
		\STATE $\textit{F}_{2}$←FindONP2(\textit{SDB}, \textit{minsup}, \textit{gap}, $\textit{F}_{1}$, $\Sigma$);
		\STATE \textit{len}←2;
		\WHILE{$\textit{F}_{len}$$\neq$NULL}
		\STATE \textit{cand}←PatternJoin($\textit{F}_{len}$, \textit{len}); // Generate all candidate patterns with length \textit{len}+1
		\FOR{each positive candidate pattern \textbf{p} in cand}
		\STATE \textit{sup}(\textbf{p}, \textit{SDB})← MatchDB(\textit{SDB}, \textbf{p}); 
		\IF{\textit{sup}(\textbf{p}, \textit{SDB})$\geq$\textit{minsup}}
		\STATE $\textit{F}_{len+1}$←$\textit{F}_{len+1}$$\cup$\textbf{p};
		\STATE Prune \textbf{p};
		\ELSE
		\STATE Prune \textbf{p} and its corresponding negative sequence patterns;
		\ENDIF
		\ENDFOR
		\STATE $\textit{F}_{len+1}$←$\textit{F}_{len+1}$$\cup$ FindFrequent(S\textit{DB}, \textit{minsup}, \textit{cand});
		\STATE \textit{len}←\textit{len}+1;
		\ENDWHILE
		\STATE return \textit{F};
	\end{algorithmic}
\end{algorithm}

ONP-Miner employs the FindONP2 algorithm to find frequent ONPs with length two, as shown in Algorithm 4. It uses the FindFrequent algorithm to discover frequent ONPs, as shown in Algorithm 5, and it adopts the PatternJoin algorithm to generate candidate patterns using the pattern join strategy, as shown in Algorithm 6.

\begin{algorithm}[h]
	\label{Algom 4}
	\caption{FindONP2} 
\hspace*{0.02in} \leftline{{\bf Input:} Sequence database \textit{SDB}, \textit{minsup}, \textit{gap}, $\textit{F}_{1}$, and $\Sigma$}\\
\hspace*{0.02in} \leftline{{\bf Output:} 2-length frequent pattern set $\textit{F}_{2}$}\\
	\begin{algorithmic}[1]
	\FOR {each character A in $\textit{F}_{1}$}
	\FOR {each character B in $\textit{F}_{1}$}
	\STATE \textit{cand}←\textit{cand} $\cup$ A$\oplus$B;
	\ENDFOR
	\ENDFOR
	\STATE $\textit{F}_{2}$←FindFrequent(\textit{SDB}, \textit{minsup}, \textit{cand});
	\FOR {each pattern \textbf{p} in $\textit{F}_{2}$}
	\FOR {each character A in $\Sigma$}
	\STATE \textbf{q}←Generate a negative candidate pattern using \textbf{p} and A;
	\STATE \textit{candn}←\textit{candn} $\cup$ \textbf{q};
	\ENDFOR
	\ENDFOR
	\STATE $\textit{F}_{2}$←$\textit{F}_{2}$ $\cup$ FindFrequent(\textit{SDB}, \textit{minsup}, \textit{candn});
	\STATE return \textit{cand};
	\end{algorithmic}
\end{algorithm}

\begin{algorithm}[h]
	\label{Algom 5}
	\caption{FindFrequent} 
	\hspace*{0.02in} \leftline{{\bf Input:} Sequence database \textit{SDB}, \textit{minsup}, and \textit{cand}}\\
	\hspace*{0.02in} \leftline{{\bf Output:} Frequent pattern set \textit{F}}\\
	\begin{algorithmic}[1]
		\FOR {each pattern \textbf{p} in \textit{cand} do}
		\STATE \textit{sup}(\textbf{p}, \textit{SDB})← MatchDB(\textit{SDB}, \textbf{p}); 
		\IF {\textit{sup}(\textbf{p}, \textit{SDB})$\geq$\textit{minsup}}
		\STATE \textit{F}← \textit{F} $\cup$ \textbf{p};
		\ENDIF
		\ENDFOR 
		\STATE return \textit{F};
	\end{algorithmic}
\end{algorithm}

\begin{algorithm}[h]
	\label{Algom 6}
	\caption{PatternJoin} 
	\hspace*{0.02in} \leftline{{\bf Input:} Frequent pattern set with length  \textit{len}, i.e., $\textit{F}_{len}$, \textit{len}}\\
	\hspace*{0.02in} \leftline{{\bf Output:} Candidate patterns with length \textit{len}+1}\\
	\begin{algorithmic}[1]
		\FOR {each pattern \textbf{p} in $\textit{F}_{len}$}
		\STATE $\textbf{p}_{suffix}$←\textit{suffix}(\textbf{p}); 
		\FOR {each pattern \textbf{q} in $\textit{F}_{len}$} 
		\STATE $\textbf{q}_{prefix}$←\textit{prefix}(\textbf{q}); 
		\IF {$\textbf{p}_{suffix}$ == $\textbf{q}_{prefix}$}
		\STATE \textbf{r}←\textbf{p}$\oplus$\textbf{q};
		\STATE \textit{cand}←\textit{cand} $\cup$ \textbf{r};
		\ENDIF
		\ENDFOR 
		\ENDFOR 
		\STATE return \textit{cand};
	\end{algorithmic}
\end{algorithm}

\textbf{Theorem 2.}
\label{Theorem2}
The time complexity of the ONP-Miner algorithm is $\textit O(l \times m \times n+l^{2})$, where \textit{l}, \textit{m}, and \textit{n} are the number of candidate patterns, the maximum length of the candidate patterns, and the length of \textit{SDB}, respectively.

\textbf{Proof}: There are two parts that affect the time complexity of the algorithm: generating candidate patterns and mining frequent patterns. Obviously, the time complexity of generating all candidate patterns is $\textit O(l^{2})$. Moreover, the time and space complexities of the MatchDB algorithm are both $\textit O(m \times n)$, since a nettree has \textit{m} levels and the number of nodes in each level does not exceed \textit{n}. The time complexity of mining frequent patterns is $\textit O(l \times m \times n)$. Therefore, the time complexity of the ONP-Miner algorithm is $\textit O(l \times m \times n+l^{2})$.

\textbf{Theorem 3.}
\label{Theorem3}
The space complexity of the ONP-Miner algorithm is $\textit O(m \times (l+n))$.

\textbf{Proof}: The space complexity of the ONP-Miner algorithm is divided into two parts: candidate pattern generation and support calculation. The space complexity of candidate patterns is $\textit O(m \times l)$, since the number of candidate patterns is \textit{l} and the number of frequent patterns does not exceed \textit{l}. The space complexity of the MatchDB algorithm is $\textit O(m \times n)$. Therefore, the space complexity of the ONP-Miner algorithm is $\textit O(m \times l+m \times n))$ = $\textit O(m \times (l+n))$.

\section{Experimental analysis}

We show the experimental datasets and the data preprocessing approaches that we used in Section 5.1. Section 5.2 introduces the principles of baseline methods. Section 5.3 evaluates the effectiveness of ONP-Miner. Section 5.4 tests the scalability of ONP-Miner. Section 5.5 compares and analyzes the mining performance of ONP-Miner. Section 5.6 describes the case study. 

All experiments are conducted on a computer with an Intel Core I5, 1.6 GHz CPU, 6 GB RAM, and the Windows 10 operating system. Our algorithm and the competitive algorithms are developed in VC++6.0, which runs as the experimental environment. The algorithms and datasets can be downloaded from https://github.com/wuc567/Pattern-Mining/tree/master/ONP-Miner.

\subsection{Benchmark datasets}

To verify the performance of ONP-Miner, we conduct experiments using data from multiple fields, including biology, transportation, finance, and game. These datasets are summarized in Table \ref{tab2}. 

\begin{table}[]
	\footnotesize
	\caption{Summary of benchmark datasets\\}
	\centering
	\label{tab2}
		\begin{tabular}{lllllll}		\hline
		Dataset & Type                   & \multicolumn{2}{l}{From}                            & Number of sequences & Maximum length & Length \\ \hline
		SDB1$ ^{\mathrm 1} $    & Protein                & \multicolumn{2}{l}{ASTRAL95\_1\_161}                & 169                 & 994            & 32,671 \\
		SDB2    & Protein                & \multicolumn{2}{l}{ASTRAL95\_1\_171}                   & 200             & 653            & 37,526 \\
		SDB3$ ^{\mathrm {2}} $   & DNA                    & \multicolumn{2}{l}{Homo sapiens AL158070}           & 1                   & 8,000          & 8,000  \\
		SDB4    & DNA                    & \multicolumn{2}{l}{Homo sapiens AL158070}           & 1                   & 10,000         & 10,000 \\
		SDB5$ ^{\mathrm{3}} $   & Activity1              & \multicolumn{2}{l}{Homo Activity}                   & 14                  & 19             & 225    \\
		SDB6    & Activity2              & \multicolumn{2}{l}{Homo Activity}                   & 21                  & 43             & 514    \\
		SDB7$ ^{\mathrm{4}} $   & Bike-sharing locations & \multicolumn{2}{l}{Bike-Chattanooga-Trip-Data}      & 37                  & 230            & 2,571  \\
		SDB8$ ^{\mathrm{5}} $   & Credit card records    & \multicolumn{2}{l}{Default of credit card clients}  & 100                 & 6              & 600    \\
		SDB9    & Diabetic symptoms      & \multicolumn{2}{l}{Diabetes\_data}                  & 314                 & 13             & 2,029  \\
		SDB10   & Traffic volume         & \multicolumn{2}{l}{Metro Interstate Traffic Volume} & 100                 & 24             & 2,400  \\ 
		SDB11$ ^{\mathrm{6}} $   & Sales         & \multicolumn{2}{l}{Video game sales with ratings} & 1                 & 100,302             & 100,302  \\ \hline
	\end{tabular}
	\begin{itemize}
	\item[1] SDB1 and SDB2 are from reference \cite{Wu 2nonNOSEP2018}; they can be downloaded from http://scop.mrc-lmb.cam.ac.uk/download.
	\item[2] SDB3 and SDB4 are from reference \cite{Wu 3close2020}; they can be downloaded from https://www.ncbi.nlm.nih.gov/nuccore/AL 158070.11.
	\item[3] SDB5 and SDB6 are from reference \cite{Wu 4con2021}; they can be downloaded from http://archive.ics.uci.edu/.
	\item[4] SDB7 can be downloaded from http://dataju.cn/.
	\item[5] SDB8, SDB9, and SDB10 can be downloaded from https://archive.ics.uci.edu/.
	\item[6] SDB11 are from reference \cite{Wu 1nonHANP2021}, they can be downloaded from http://dataju.cn.
\end{itemize}
\end{table}

The Activity dataset is divided into Activity1 and Activity2, since we want to determine the influence of the sequence length on mining results. Therefore, sequences with a length of less than 20 are classified as Activity1, and sequences with a length that is no less than 20 are classified as Activity2. Moreover, datasets SDB7 to SDB10 are numerical datasets. To mine frequent ONPs on these numerical datasets, we convert the datasets into character datasets. The processing methods are as follows.

1. For SDB7, we select the 30-day shared bicycle starting point location data and convert the numerical location label data into a character sequence. The locations are divided into nine regions according to their values, which are represented by `a', `b', `c', `d', `e', `f', `g', `h', and `i'. We assign a character to each region. 

2. For SDB8, we represent the repayment of credit cards. Timely repayment is represented by `a', repayment delayed for one day by `b', repayment delayed for two days by `c', and so on. We get the character set $\Sigma$ = \{`a', `b', `c', `d', `e', `f', `g', `h', `i', `j'\}.

3. For SDB9, the obtained diabetic symptom data was converted into a character sequence. Diabetic symptoms are represented by characters; for example, `A' = polyuria, `B' = polydipsia, and `C' = sudden weight loss. We get the character set $\Sigma$ = \{`A', `B', `C', `D', `E', `F', `G', `H', `I', `J', `K', `L', `M'\}.

4. For SDB10, the traffic volume data for 100 days are selected. We count the traffic volume per hour. The maximum traffic volume per hour is less than 7000 . We use seven characters, `a', `b', `c', `d', `e', `f', and `g', to represent the volume; they represent traffic volumes between 0 and 1000, between 1001 and 2000, between 2001 and 3000, between 3001 and 4000, between 4001 and 5000, between 5001 and 6000, and between 6001 and 7000, respectively. 

\subsection{Baseline methods}

We evaluate the effectiveness of ONP-Miner in Section 5.3. ONP-Miner consists of support calculation and candidate pattern generation. We evaluate the effectiveness of these two parts. Therefore, we propose comparing ONP-Miner with the following competitive algorithms. The principles of these algorithms follow.

1. SRMP-N: To assess the effectiveness of MatchDB, SRMP-N is proposed. It employs the SRMP algorithm \cite{Yu  SRMP2021} to calculate the support to mine ONPs and employs the pattern join and pruning strategies to generate candidate patterns.

2. ONP-BFS and ONP-DFS: To validate the effectiveness of the pattern join strategy, ONP-BFS and ONP-DFS are proposed. They adopt the breadth-first and depth-first strategies, respectively, to generate candidate patterns, and they use MatchDB to calculate the support.

3. ONP-ALL: To evaluate the effectiveness of the pruning strategy, ONP-ALL is proposed. It uses the pattern join strategy to generate candidate patterns and uses MatchDB to calculate the support.

4. e-NSP \cite{Cao  6neg2016ensp} and e-RNSP \cite{Dong  7neg2020ernsp}: To further validate the mining performance of the ONP-Miner algorithm on negative sequential patterns, we select these two state-of-the-art algorithms as competitive algorithms. 

\subsection{Effectiveness}

\begin{figure}[]
	\centering
	\includegraphics[width=0.45\linewidth]{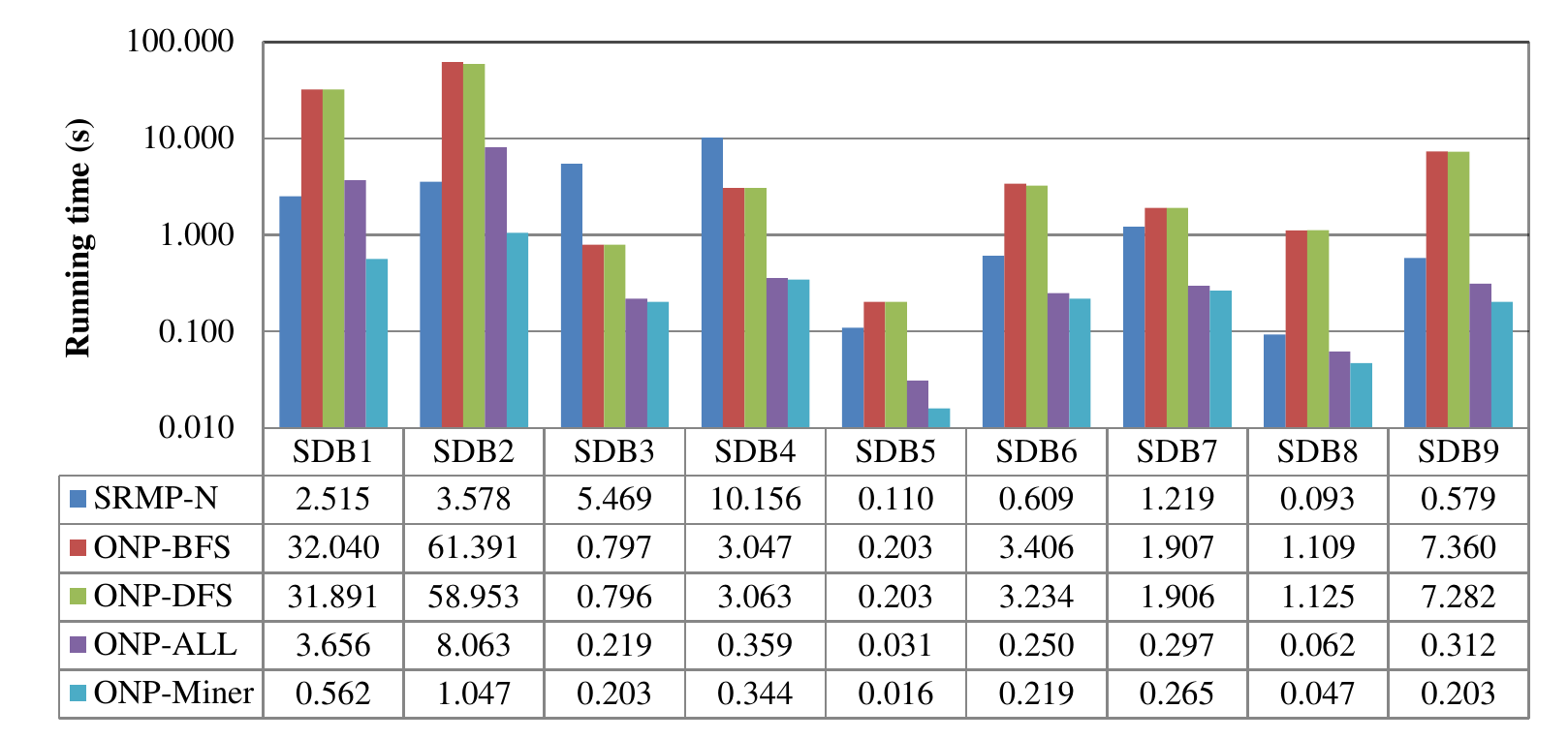}
	\caption{Comparison of the running time.}
	\label{fig:figure5}
\end{figure}

\begin{figure}[]
	\centering
	\includegraphics[width=0.45\linewidth]{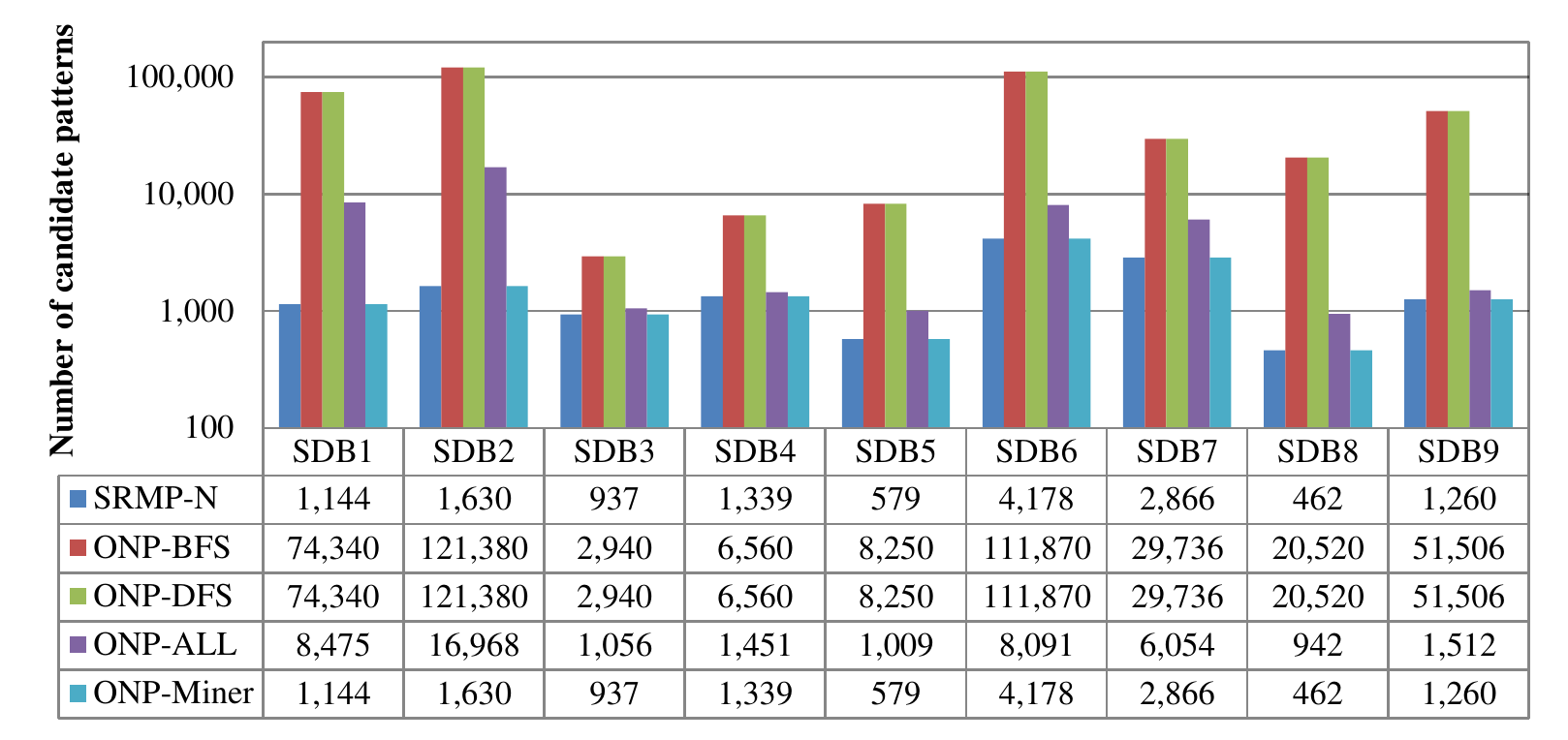}
	\caption{Comparison of the number of candidate patterns.}
	\label{fig:figure6}
\end{figure}

In this section, we will assess the effectiveness of ONP-Miner using different strategies. Four competitive algorithms are selected: SRMP-N, ONP-BFS, ONP-DFS, and ONP-ALL. We use nine databases, SDB1–SDB9, to conduct the experiments. Since the lengths of these datasets are different, to mine a moderate number of patterns, we use different parameters for different datasets: for SDB1–SDB4, \textit{gap} = [0,3] and \textit{minsup} = 500; for SDB5–SDB6, \textit{gap} = [0,2] and \textit{minsup} = 15; for SDB7, \textit{gap} = [0,2] and \textit{minsup} = 60; for SDB8, \textit{gap} = [0,2] and \textit{minsup} = 20; and for SDB9, \textit{gap} = [0,2] and \textit{minsup} = 80. All of the algorithms mine 157, 269, 143, 401, 65, 975, 151, 219, and 270 ONPs from SDB1–SDB9, respectively. The comparisons of the running time and the number of candidate patterns are shown in Figs. ~\ref{fig:figure5}  and ~\ref{fig:figure6}, respectively.


The results give rise to the following observations. 

1. The MatchDB algorithm is significantly efficient in calculating the support. According to Figs. ~\ref{fig:figure5} and \ref{fig:figure6}, SRMP-N and ONP-Miner generate the same number of candidate patterns, but the running time of ONP-Miner is less than that of SRMP-N. For example, for SDB1, the two algorithms both generate 1,144 candidate patterns and mine 157 ONPs. However, the running time of SRMP-N and ONP-Miner are 2.094 and 0.625 s, respectively. This phenomenon can be found on all of the other datasets. We know that the difference between SRMP-N and ONP-Miner is that the two algorithms adopt different methods to calculate the support. ONP-Miner runs faster than SRMP-N, which indicates that MatchDB is more effective than SRMP, since ONP-Miner employs the MatchDB algorithm to calculate the support, which can reduce the running time by avoiding the creation of invalid nodes. Thus, ONP-Miner is better than SRMP-N.

2. The pattern join and pruning strategies are more efficient than the other strategies. According to Fig.~\ref{fig:figure5}, the running time of ONP-Miner is less than that of ONP-BFS, ONP-DFS, and ONP-ALL. For example, on SDB9, the running time of ONP-BFS, ONP-DFS, and ONP-ALL is 7.360, 7.282, and 0.469 s, respectively, while that of ONP-Miner is 0.297 s. The reason for this is that ONP-Miner calculates fewer candidate patterns. According to Fig.~\ref{fig:figure6}, for SDB9,ONP-Miner only calculates 1,274 candidate patterns, while ONP-BFS, ONP-DFS, and ONP-ALL calculate 51,506, 51,506, and 2,878 candidate patterns, respectively. We know that ONP-Miner employs the pattern join and pruning strategies to generate candidate patterns, while ONP-BFS, ONP-DFS, and ONP-ALL employ the breadth-first, depth-first, and pattern join strategies, respectively. Thus, the results show that the pattern join and pruning strategies are more efficient than the breadth-first, depth-first, and pattern join strategies. Hence, ONP-Miner outperforms ONP-BFS, ONP-DFS, and ONP-ALL.

Therefore, ONP-Miner is superior to all competitive algorithms. 

\subsection{Scalability}

\begin{figure}[]
	\centering
	\includegraphics[width=0.45\linewidth]{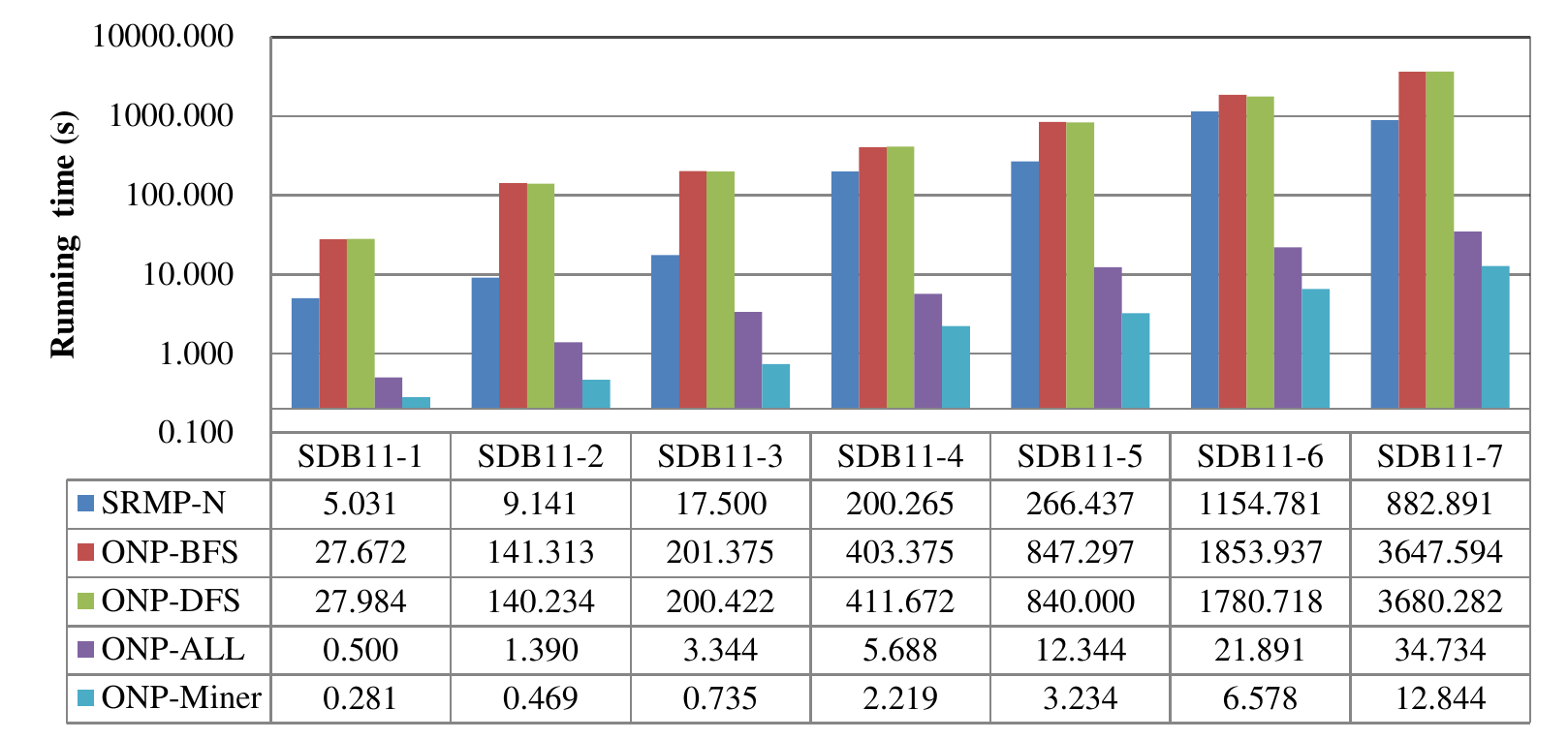}
	\caption{Comparison of the running time of scalability. }
	\label{fig:figure7}
\end{figure}

\begin{figure}[]
	\centering
	\includegraphics[width=0.45\linewidth]{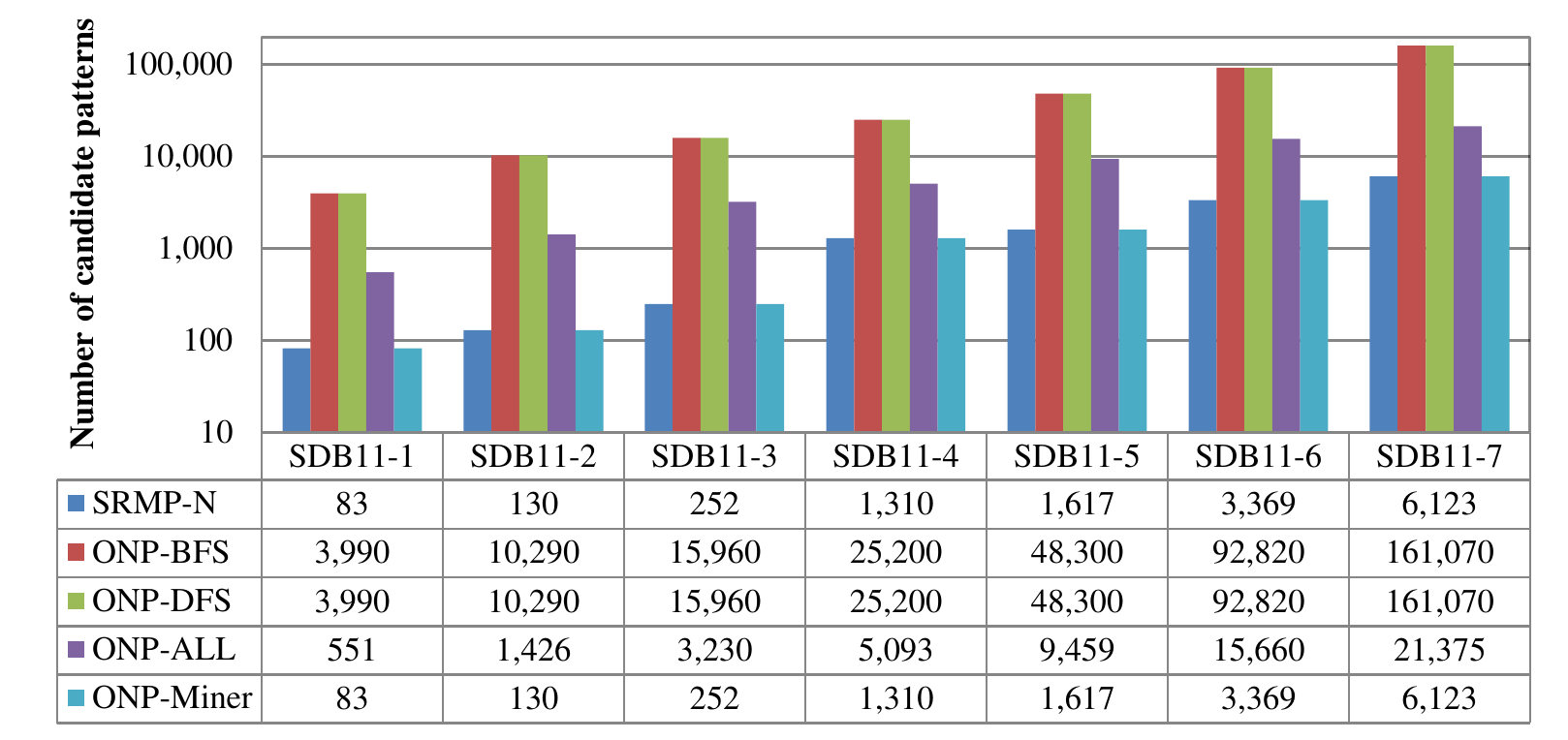}
	\caption{Comparison of the number of candidate patterns of scalability.}
	\label{fig:figure8}
\end{figure}

In this section, we will test the scalability of ONP-Miner under different dataset lengths. We carry out experiments on SDB11, and the subsequences of SDB11 with length of 40,000, 50,000, 60,000, 70,000, 80,000, and 90,000 are named SDB11-1 to SDB11-6, respectively, and SDB11 is named SDB11-7. The experimental parameters are \textit{gap} = [0,4] and \textit{minsup} = 3000. All of the algorithms mine 5, 35, 62, 106, 216, 428, and 753 ONPs from SDB11-1 to SDB11-7, respectively. The comparisons of the running time and the number of candidate patterns are shown in Figs. ~\ref{fig:figure7} and ~\ref{fig:figure8}, respectively. 


The results give rise to the following observations. 

1. With the increase of the length of dataset, the number of frequent patterns is increasing, and the running time and the number of candidate patterns are also increasing. According to Figs. ~\ref{fig:figure7} and ~\ref{fig:figure8}, when the parameters of the algorithms remain unchanged, the number of frequent patterns, running time and the number of candidate patterns increase with the length of the datasets. For example, when the length of dataset is 40,000, ONP-Miner mines 5 frequent patterns from SDB11-1, runs 0.281 s, and generates 83 candidate patterns, while when the length of dataset is 90,000, it mines 428 frequent patterns from SDB11-6, runs 6.578 s, and generates 3,369 candidate patterns. This phenomenon can also be found in other competitive algorithms.

2. With the increase of the length of dataset, ONP-Miner has better mining performance. According to  Figs. ~\ref{fig:figure7} and ~\ref{fig:figure8}, when mining the same number of frequent patterns, ONP-Miner has the shortest running time and the least number of candidate patterns. For example, in SDB11-7 with the longest data length, ONP-Miner mines 753 frequent patterns the same as other competitive algorithms. However, ONP-Miner only runs 12.844 s and generates 6123 candidate patterns, while SRMP-N, ONP-BFS, ONP-DFS, and ONP-ALL run 882.891, 3647.594, 3680.282, and 34.734 s and generate 6123, 161070, 161070, and 21375 candidate patterns, respectively.

Therefore, ONP-Miner has strong scalability, and the mining performance will not degrade with the increase of dataset length.

\subsection{Mining performance}

In this section, we report the mining performance of the ONP-Miner algorithm on negative sequential patterns, and we select two most similar algorithms as the competitive algorithms: e-NSP \cite{Cao  6neg2016ensp} and e-RNSP \cite{Dong  7neg2020ernsp}. Before the experiment, we clarify the difference between our algorithm and the competitive algorithms.  e-NSP \cite{Cao  6neg2016ensp} and e-RNSP \cite{Dong  7neg2020ernsp} are two similar algorithms, but the problems they solved are different from ours. An illustrative example is as follows. Suppose pattern  `abc' is a frequent positive pattern. According to e-NSP and e-RNSP, `a¬bc' is a negative candidate pattern based on pattern `abc'. More importantly, pattern `a¬bc' occurs in a sequence means that the sequence has `ac' but does not have `abc'. Therefore, \textit{sup}(a¬bc)= \textit{sup}(ac)- \textit{sup}(abc) \cite{Cao  6neg2016ensp, Dong  7neg2020ernsp}. However, in our problem, we consider the gap constraints [\textit{M}, \textit{N}]. The negative candidate pattern a[\textit{M}, \textit{N}]¬bc is generated based on frequent pattern `a[\textit{M}, \textit{N}]c', rather than pattern `a[M, N]b[\textit{M}, \textit{N}]c'. Therefore, the problems e-NSP \cite{Cao  6neg2016ensp} and e-RNSP \cite{Dong  7neg2020ernsp} solved are far different from ours. To clearly report the performance of ONP-Miner, we conduct two experiments. In the first, we set different parameters to make the three algorithms mine similar numbers of frequent patterns. In the second, we set different parameters to make the three algorithms mine a similar number of frequent positive patterns. We know that e-NSP can only be conducted on multiple sequences, since e-NSP does not consider the repetition of patterns in a sequence. Hence, in the two experiments, we select multiple sequences datasets SDB5-SDB9.

\subsubsection{Comparative experiment of similar number of frequent patterns}
\

In this experiment, we set different minsup values to make the three algorithms mine similar numbers of frequent patterns. Moreover, ONP-Miner sets the gap constraint from 0 to the maximum value of the sequence, since e-RNSP does not consider gap constraints. For example, on SDB5, ONP-Miner sets $gap$ = [0,19], where 19 is the maximum sequence length. The parameters of each algorithm are shown in Table \ref{tab3}.

\begin{table}[]
	\footnotesize
	\caption{Parameters of each algorithm for mining similar number of frequent patterns}
	\centering
	\label{tab3}
	\begin{tabular}{lllllll}
		\hline
		Dataset & e-NSP(\textit{minsup})                   & \multicolumn{2}{l}{e-RNSP(\textit{minsup})}                         & ONP-Miner(\textit{minsup}) & ONP-Miner(\textit{gap}) \\ \hline
SDB5   & 7    & \multicolumn{2}{l}{7}          & 12       & [0.19]             \\
SDB6    & 16  & \multicolumn{2}{l}{16}         & 30       & [0.43]             \\
SDB7   & 24   & \multicolumn{2}{l}{53}         & 109      & [0,230]           \\
SDB8    & 10     & \multicolumn{2}{l}{10}      & 17       & [0,6]         \\
SDB9   & 50      & \multicolumn{2}{l}{50}      & 102      & [0,13]\\ \hline
	\end{tabular}
\end{table}

\begin{figure}
	\centering
	\includegraphics[width=0.4\linewidth]{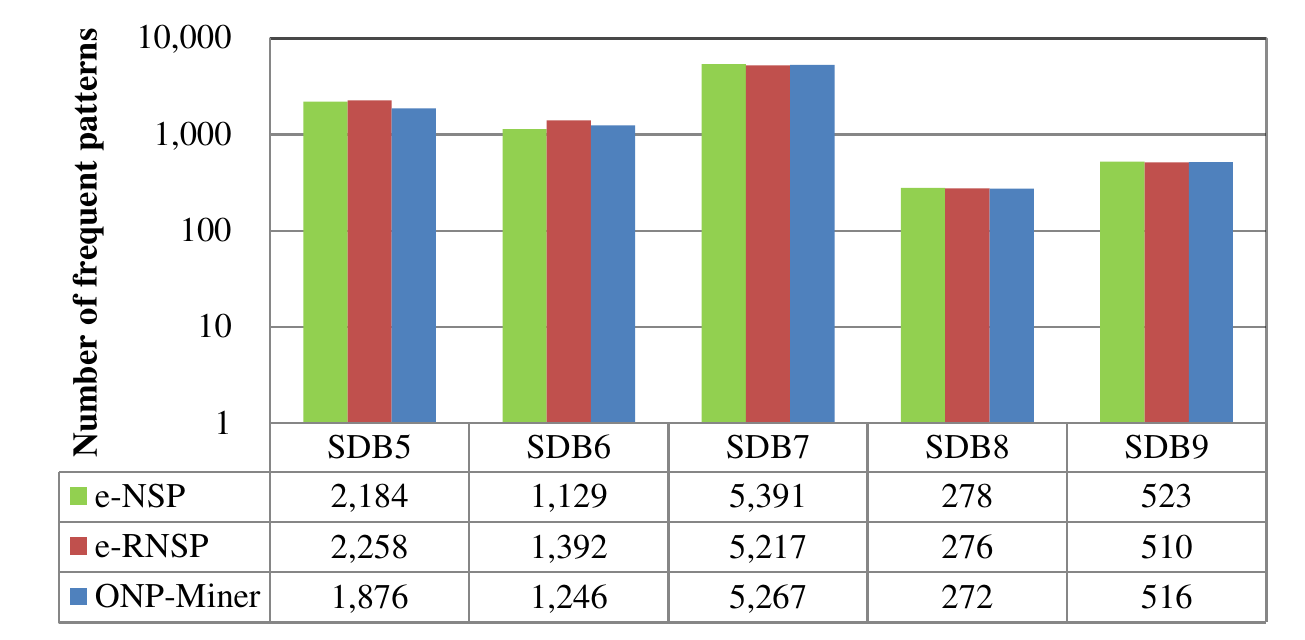}
	\caption{Comparison of the number of frequent patterns with similar number of frequent patterns. }
	\label{fig:figure9}
\end{figure}

\begin{figure}
	\centering
	\includegraphics[width=0.4\linewidth]{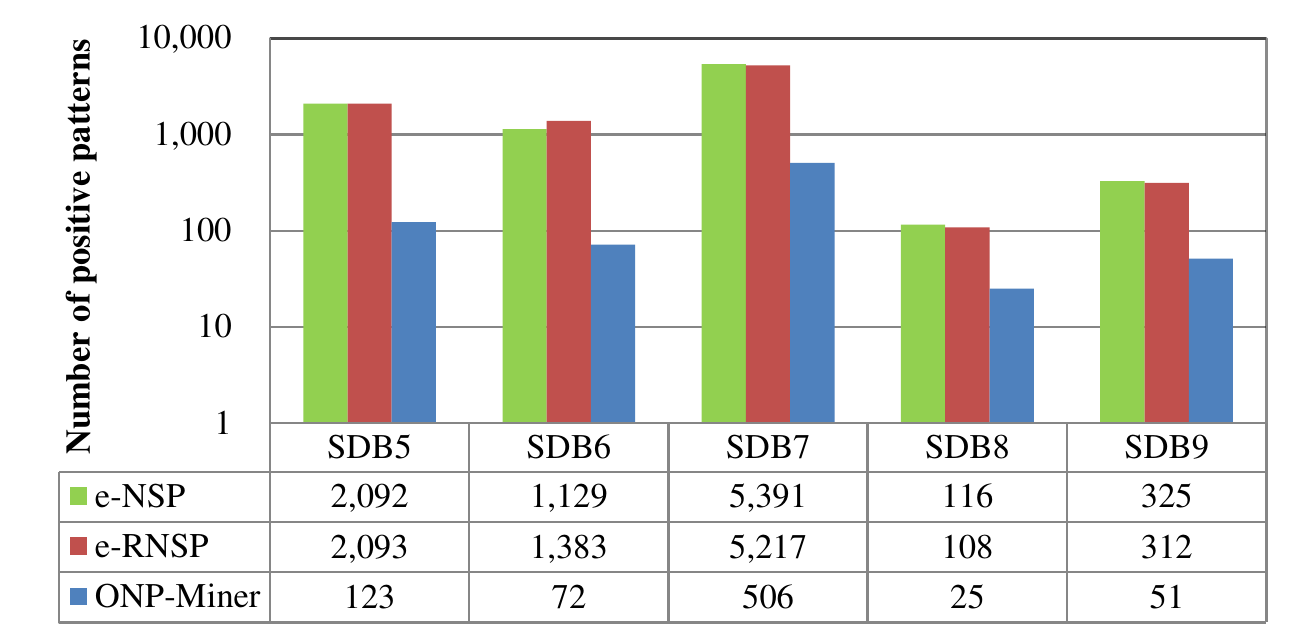}
	\caption{Comparison of the number of positive patterns with similar number of frequent patterns.}
	\label{fig:figure10}
\end{figure}

\begin{figure}
	\centering
	\includegraphics[width=0.4\linewidth]{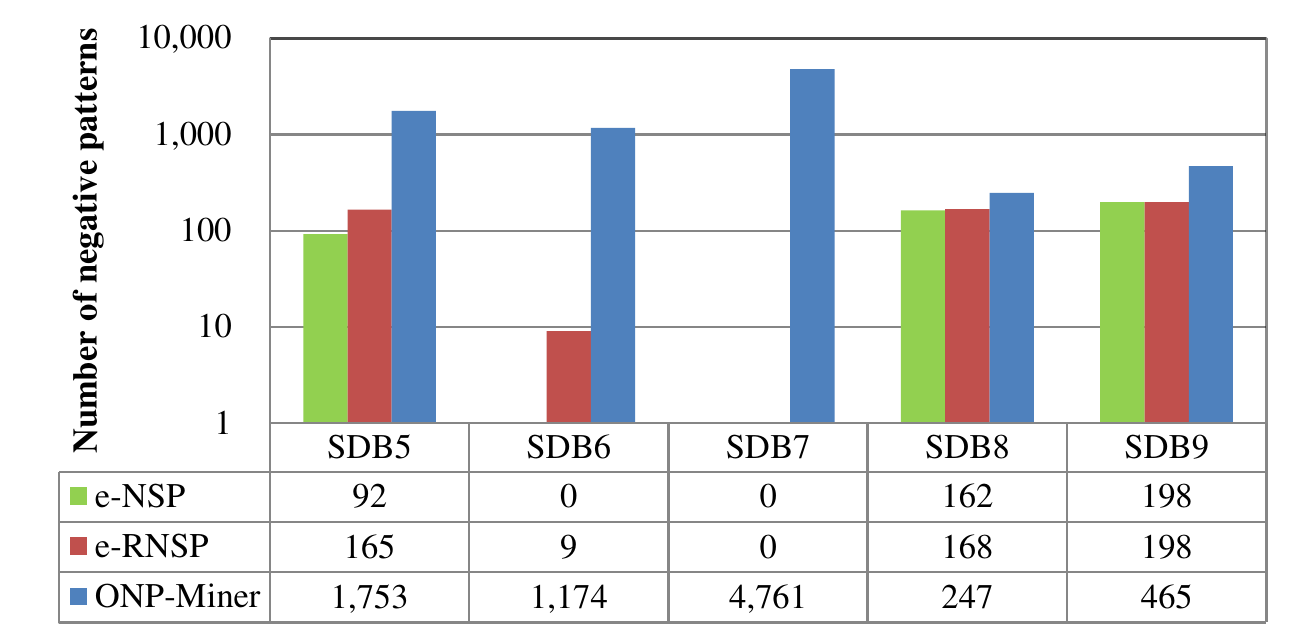}
	\caption{Comparison of the number of negative patterns with similar number of frequent patterns.}
	\label{fig:figure11}
\end{figure}

\begin{figure}
	\centering
	\includegraphics[width=0.4\linewidth]{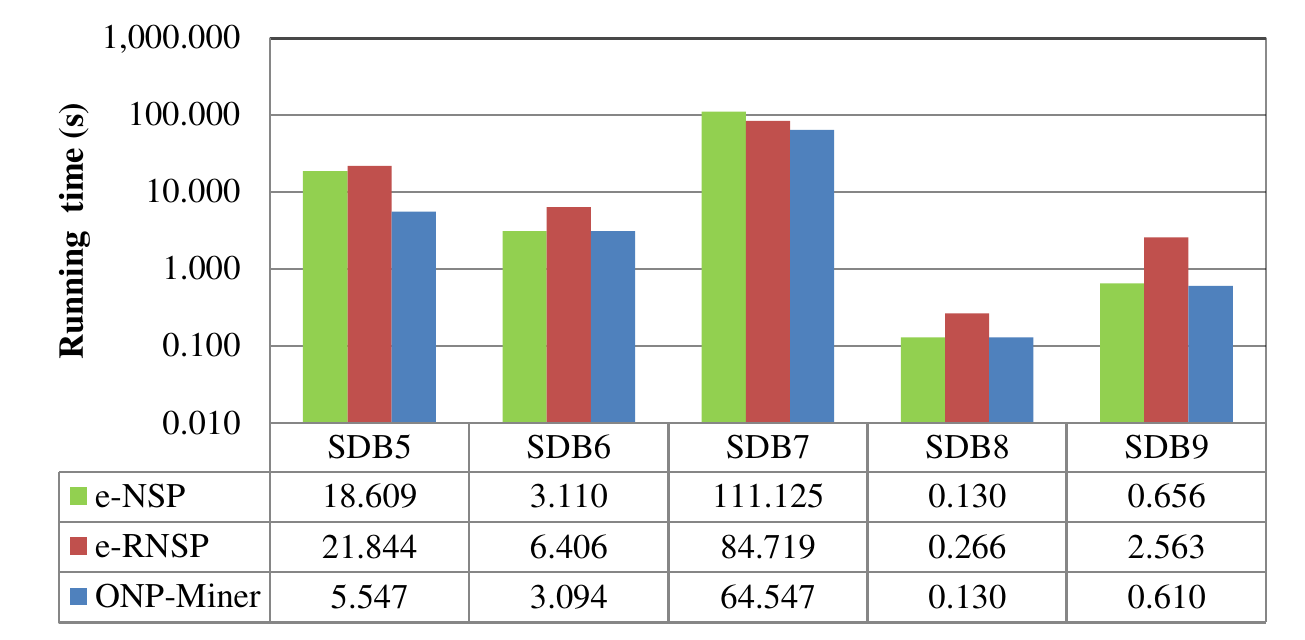}
	\caption{Comparison of the running time with similar number of frequent patterns.}
	\label{fig:figure12}
\end{figure}

The comparisons of the number of frequent patterns, positive patterns, negative patterns, and running time are shown in Figs. ~\ref{fig:figure9} -~\ref{fig:figure12}, respectively.

The results give rise to the following observations.

1. ONP-Miner can mine similar number of frequent patterns faster than e-NSP and e-RNSP. For example, e-NSP, e-RNSP, and ONP-Miner discover 523, 510, and 516 frequent patterns in SDB9, respectively. However, the running time of e-NSP, e-RNSP, and ONP-Miner is 0.656, 2.563, and 0.610 s, respectively. This phenomenon can be found on all the other datasets. The results indicate that ONP-Miner is more effective in mining negative patterns, since it employs the MatchDB algorithm to calculate the support and the pattern join with pruning method to generate candidate patterns. As mentioned in the previous section, we confirm that MatchDB and the pattern join strategy with pruning are effective methods for calculating the support and generating candidate patterns, respectively.

2. ONP-Miner can find negative sequential patterns in all kinds of datasets, while e-NSP and e-RNSP cannot find negative sequential patterns in the datasets with long sequences and small character set. For example, as shown in  Fig. ~\ref{fig:figure11}, ONP-Miner mines 1,174 frequent negative patterns from SDB6, while e-RNSP can only mine 9 frequent negative patterns, and e-NSP cannot mine any frequent negative patterns. Moreover, neither e-NSP nor e-RNSP can find negative patterns in SDB7, while ONP-Miner mines 4,761 negative patterns. From Table ~\ref{tab2}, we know that the maximum sequence lengths of SDB6 and SDB7 are 43 and 230, respectively. However, e-NSP and e-RNSP can find negative patterns in SDB5, SDB8, and SDB9, whose maximum sequence lengths are 19, 6, and 13, respectively. The results indicate that e-NSP and e-RNSP cannot find negative sequential patterns in the datasets with long sequences and small character set. We take SDB7 as an example. The maximum sequence length of SDB7 is 230 and there are only nine characters, which means that the nine characters exist in almost every sequence. Therefore, e-NSP and e-RNSP cannot find the negative patterns. However, ONP-Miner considers both the repetition of patterns and gap constraints, which are not limited by the sequence length and the number of characters. Thus, ONP-Miner can find the negative patterns.

3. ONP-Miner can discover more negative patterns than other competitive algorithms. For example, for SDB9 in Figs. ~\ref{fig:figure9} -~\ref{fig:figure12}, e-NSP, e-RNSP, and ONP-Miner discover 523, 510, and 516 frequent patterns, respectively. Among them, e-NSP finds 198 negative patterns and 325 positive patterns, e-RNSP mines 198 negative patterns and 312 positive patterns, while ONP-Miner discovers 465 negative patterns and 51 positive patterns. This phenomenon can be found on all other datasets. The reason for this is described as follows. Suppose that pattern  `abc' is a frequent positive pattern, and the character set is $\Sigma$ = \{a, b, c\}. For e-NSP and e-RNSP, `a¬bc' can be a candidate negative pattern, while for ONP-Miner, there are three candidate negative patterns, `a¬abc', `a¬bbc', and `a¬cbc'. Moreover, `a¬b¬c' is an illegal candidate negative pattern for e-NSP and e-RNSP. However, `a¬ab¬ac' can be a candidate negative pattern according to ONP-Miner. Thus, ONP-Miner has more candidate negative patterns than e-NSP and e-RNSP. Hence, ONP-Miner can discover more negative patterns than e-NSP and e-RNSP. To further verify that ONP-Miner can discover more negative patterns, we set parameters to enable e-NSP, e-RNSP, and ONP-Miner to mine similar number of frequent positive patterns, and then compare the number of frequent negative patterns.

\subsubsection{Comparative experiment of similar number of frequent positive patterns} 
\
                
In this experiment, we also select SDB5-SDB9, and set different minsup values to make the three algorithms mine similar number of frequent positive patterns. The gap constraints of ONP-Miner are the same as Table \ref{tab3}, and the thresholds are shown in Table \ref{tab4}.

\begin{table}[]
	\footnotesize
	\caption{ Parameters of each algorithm for mining similar number of frequent positive patterns}
	\centering
	\label{tab4}
	\begin{tabular}{lllllll}
		\hline
		Dataset & e-NSP(\textit{minsup})    & \multicolumn{2}{l}{e-RNSP(\textit{minsup})}                            & ONP-Miner(\textit{minsup}) \\ \hline
SDB5   & 7               & \multicolumn{2}{l}{7}            & 7    \\
SDB6   & 15              & \multicolumn{2}{l}{15}           & 16   \\
SDB7   & 28              & \multicolumn{2}{l}{62}           & 102  \\
SDB8   & 20              & \multicolumn{2}{l}{20}           & 20   \\
SDB9   & 100             & \multicolumn{2}{l}{100}          & 100  \\ \hline
	\end{tabular}
\end{table}

\begin{figure}
	\centering
	\includegraphics[width=0.4\linewidth]{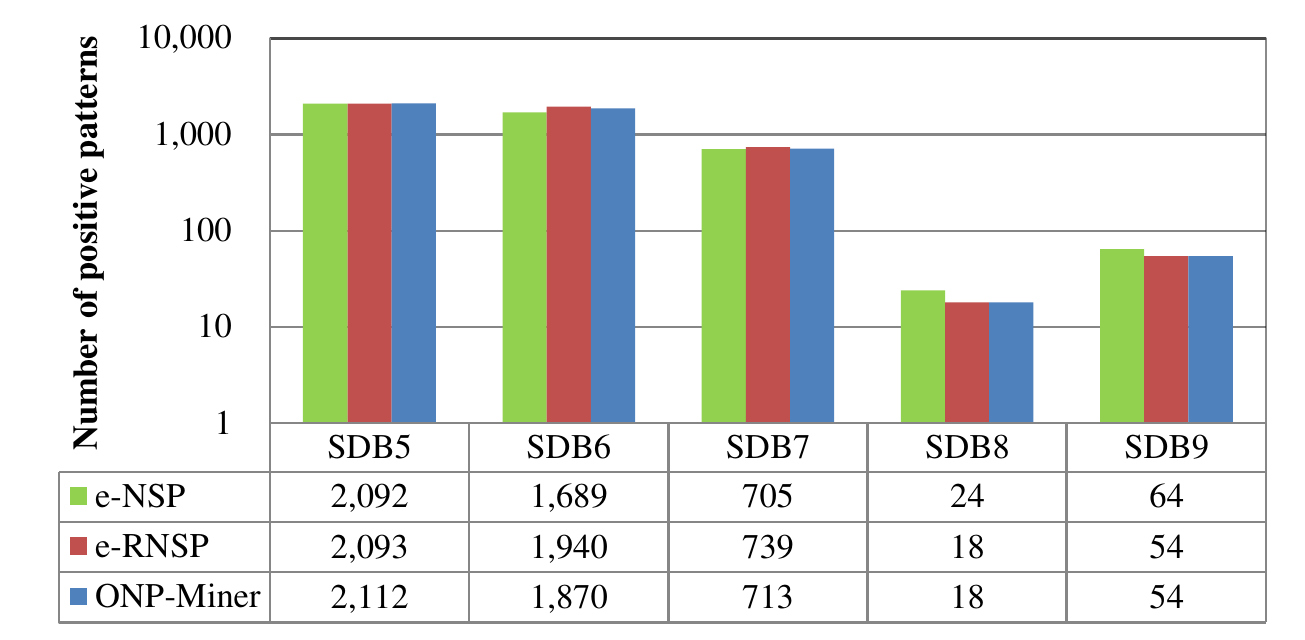}
	\caption{Comparison of the number of positive patterns with similar number of frequent positive patterns.}
	\label{fig:figure13}
\end{figure}

\begin{figure}
	\centering
	\includegraphics[width=0.4\linewidth]{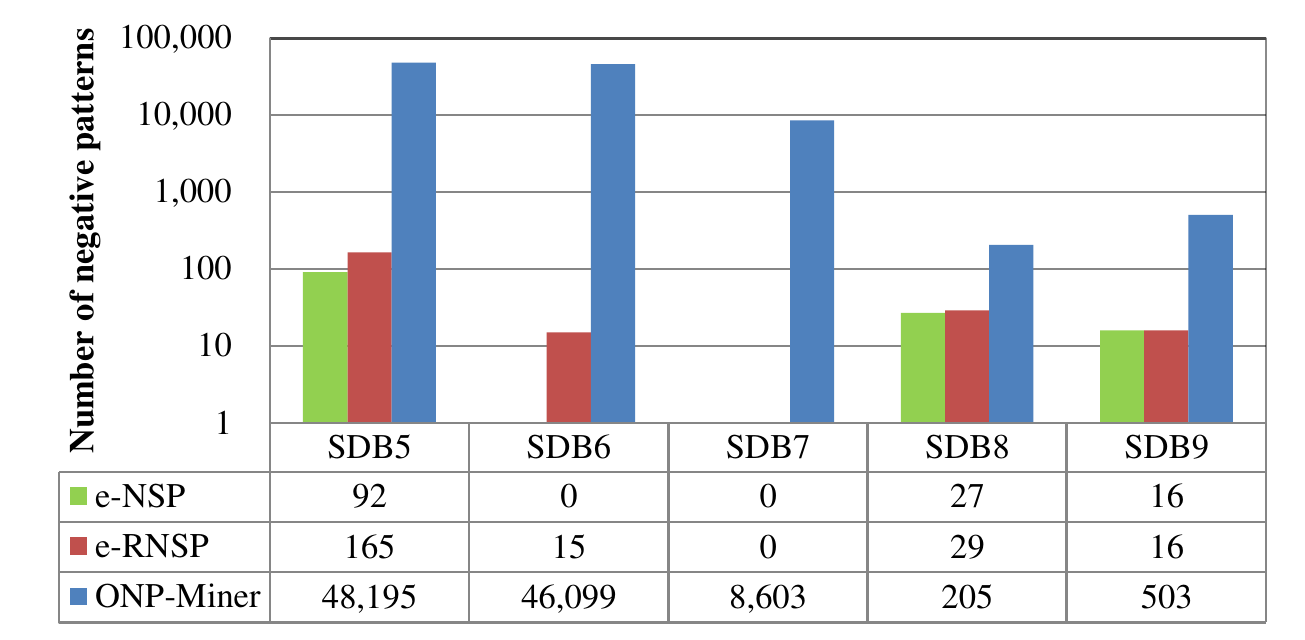}
	\caption{ Comparison of the number of negative patterns  with similar number of frequent positive patterns.}
	\label{fig:figure14}
\end{figure}

\begin{figure}
	\centering
	\includegraphics[width=0.4\linewidth]{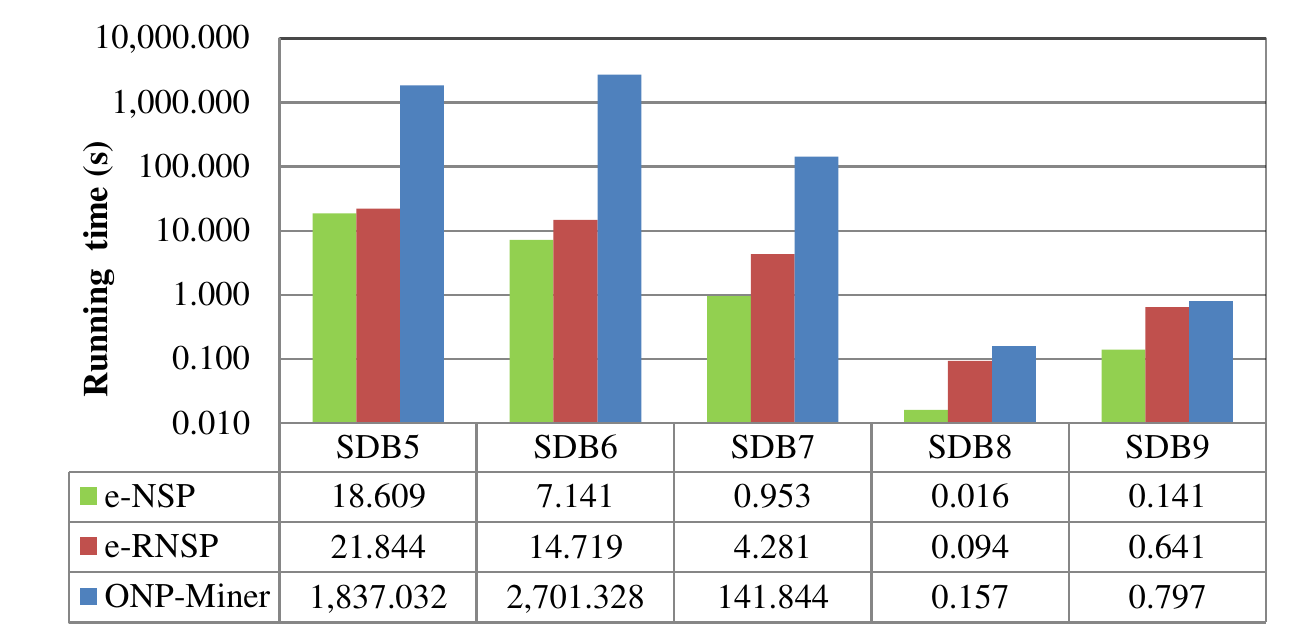}
	\caption{ Comparison of the running time  with similar number of frequent positive patterns.}
	\label{fig:figure15}
\end{figure}

\begin{figure}
	\centering
	\includegraphics[width=0.4\linewidth]{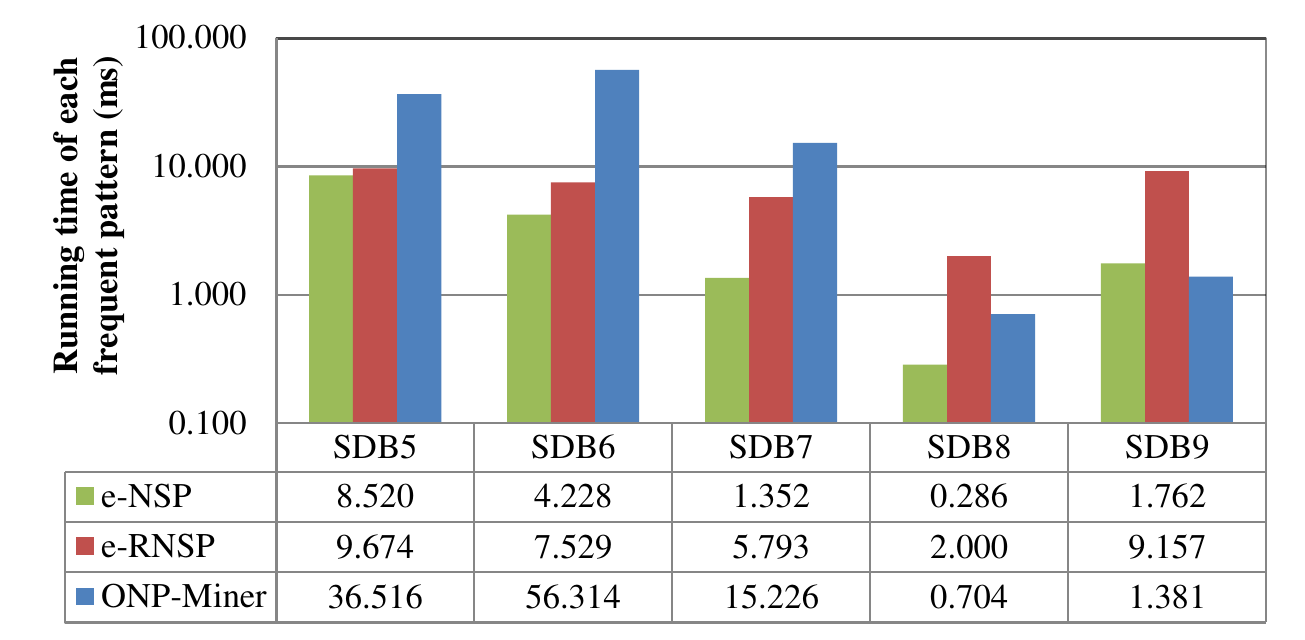}
	\caption{Comparison of the average running time of each frequent pattern  with similar number of frequent positive patterns.}
	\label{fig:figure16}
\end{figure}

The comparisons of the number of positive patterns, negative patterns, running time, and average running time of each pattern are shown in Figs. ~\ref{fig:figure13} -~\ref{fig:figure16}, respectively.

According to Figs. ~\ref{fig:figure13} and  ~\ref{fig:figure14}, ONP-Miner algorithm can mine more frequent negative patterns than e-NSP and e-RNSP. For example, on SDB9, e-NSP mines 64 frequent positive patterns and 16 frequent negative patterns, and e-RNSP mines 54 frequent positive patterns and 16 frequent negative patterns, while ONP-Miner mines 54 frequent positive patterns and discovers 503 frequent negative patterns. Therefore, the results validate that ONP-Miner can mine more frequent negative patterns. The reason is analyzed in the previous subsection.

It is worth noting that Figs. ~\ref{fig:figure15} and ~\ref{fig:figure16}  show that ONP-Miner has shortcomings in the running time and the average running time of each pattern. For example, on SDB5, the running time of ONP-Miner is 1837.032s, and the average running time of each pattern is 36.516 ms, which are worse than e-NSP and e-RNSP. This result indicates that a more effective ONP-Miner algorithm is worth investigating.

\subsection{Case study}

\begin{figure*}[]
	\centering
	\subfigure[Frequent positive patterns]{
		\centering
		\includegraphics[width=0.35\linewidth]{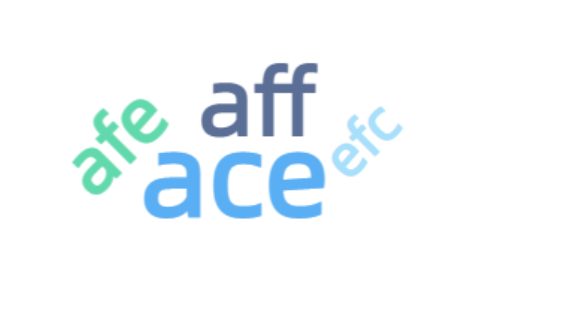}
		\label{fig:figure17a}
	}%
	\subfigure[Frequent negative patterns]{
		\centering
		\includegraphics[width=0.45\linewidth]{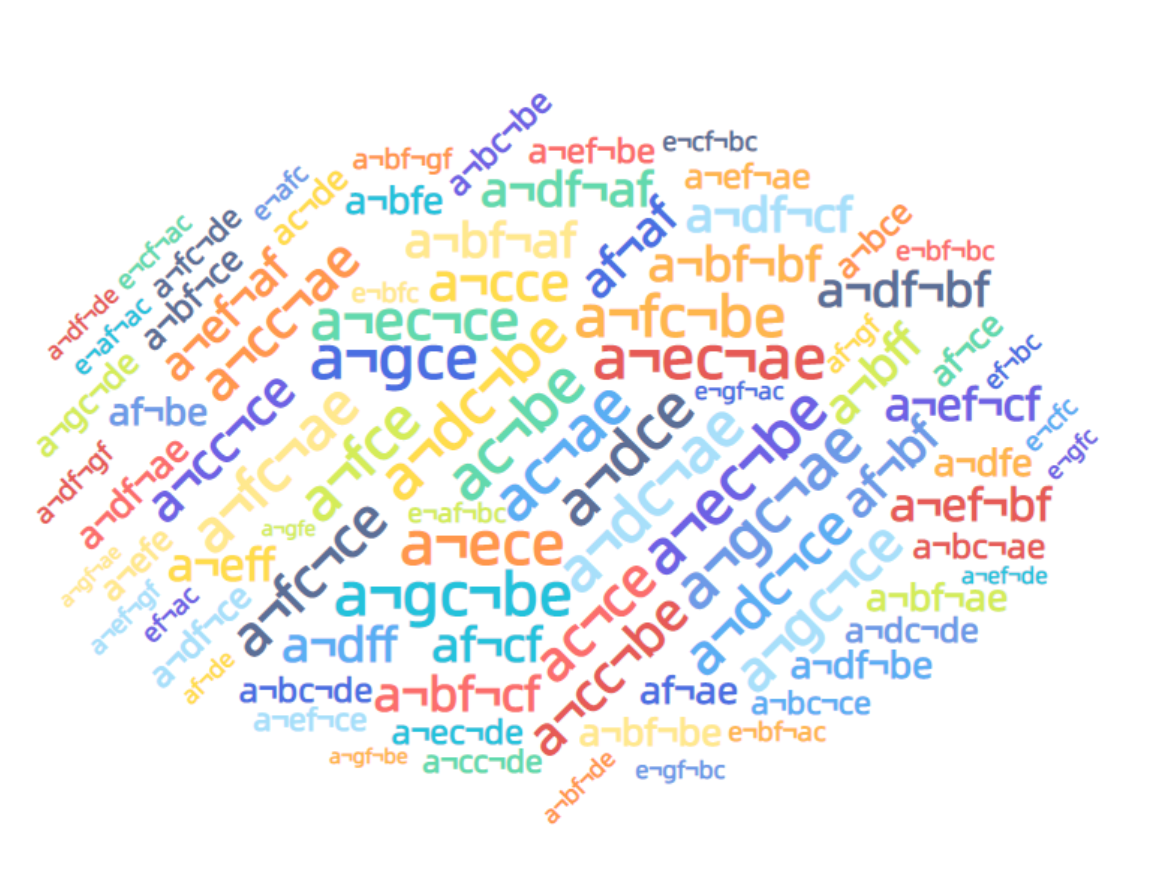}
		\label{fig:figure17b}
	}%
	\centering
	\caption{Mining results of frequent patterns with length three.}
	\label{fig:figure17}
\end{figure*}

In the experiment, we set \textit{gap} = [4,4] to measure the change in the traffic volume every four hours. Thus, a pattern with length of three, such as pattern a[4,4]¬bc[4,4]¬de, can describes the traffic volumes of 11 hours, since each character represents the traffic volumes of one hour. Therefore, a specific pattern can occur at most twice a day. We know that SDB10 contains 100 days of traffic volumes. We set \textit{minsup} = 50, i.e. a pattern must occur at least 50 times. If a pattern occurs once a day, then \textit{minsup} = 50 indicates that a pattern happens in 50 days. If a pattern occurs twice a day, then \textit{minsup} = 50 indicates that a pattern happens in 25 days. We select \textit{minsup} = 50 to indicate that a pattern is a frequent pattern. We can discover 178 frequent patterns, of which 85 are patterns with length two and 93 are patterns with length three. Now, we will analyze the 93 patterns. There are four positive patterns and 89 negative patterns. The word cloud map of these patterns is shown in Fig.\ref{fig:figure17}.

To further analyze the negative patterns in Fig. \ref{fig:figure17b},  Table \ref{tab5} shows the statistics of the negative patterns in each time period.

\begin{table}[]
	\footnotesize
	\caption{Statistics of negative patterns in each time period}
	\centering
		\label{tab5}
	\begin{threeparttable}

	\begin{tabular}{lllll}
		\hline
The first hour & The next four hours & The sixth hour & The next four hours                                                                        & The 11$^{\mathrm{th}}$ hour \\ \hline

a(75)          & \begin{tabular}[c]{@{}l@{}}¬b(15), ¬c(5), ¬d(15), \\ ¬e(15), ¬f(5), ¬g(8), \\ Null(12)\end{tabular} & c(34), f(41)   & \begin{tabular}[c]{@{}l@{}}¬a(16), ¬b(16), ¬c(15),\\  ¬d(11), ¬g(4), Null(13)\end{tabular} & e(56), f(19)  \\

e(14)          & \begin{tabular}[c]{@{}l@{}}¬a(3), ¬b(3), ¬c(3),\\  ¬g(3), Null(2)\end{tabular}                      & f(14)          & ¬a(5), ¬b(5), Null(4)                                                                      & c(14)         \\ \hline
	\end{tabular}
\begin{tablenotes}
		\footnotesize
		\item[] Note: The number in brackets indicates the frequency of the corresponding character.
\end{tablenotes}
\end{threeparttable}
\end{table}

According to Section 5.1, character `a' represents a traffic volume of 0–1000, which is the minimal level of traffic volume. Characters `b', `c', and `d' belong to the medium traffic level. Characters `e' and `f' belong to the above-medium traffic level. Character `g' belongs to the maximal traffic level. The results give rise to the following observations.

1. Not only is the number of frequent positive patterns small, but also it is difficult to get interesting information. From Fig. \ref{fig:figure17a},  it is clear that there are only four frequent positive patterns. Obviously, these patterns are meaningless, since we can only know the frequent patterns from the traffic flow sequence formed in the first hour, the sixth hour, and the eleventh hour. Hence, it is difficult to use these patterns to guide practice.

2. Not only is the number of frequent negative patterns large, but also it is easy to obtain interesting information. There are 89 frequent negative patterns in Fig. \ref{fig:figure17b} .  Moreover, there are two cases according to Table~\ref{tab5}.

Case 1: When the traffic volume in the first hour is `a', this indicates that the traffic volume in this period is very small. In the next four hours, the traffic volume can be ¬b, ¬c, ¬d, ¬e, ¬f, ¬g, and Null, indicating that the traffic volume has not increased rapidly in these four hours. In the sixth hour, the traffic volume can be `c' or `f'. Thus, it belongs to the medium or above-medium levels. In the next four hours, the traffic volume can be ¬a, ¬b, ¬c, ¬d, ¬g, and Null, indicating that the traffic volume in these four hours has not decreased or peaked, but it is still above the medium level. In the 11$^{\mathrm{th}}$ hour, the traffic volume can be `e' or `f', so it belongs to the above-medium level. These patterns indicate that when the initial traffic volume is very small, the traffic volume begins to increase gradually, reaches the medium or above-medium level after four hours, and then continues to remain within this level.

Case 2: When the traffic volume is `e' in the first hour, it belongs to the above-medium level. In the next four hours, the traffic volume can be ¬a, ¬b, ¬c, ¬g, and Null, indicating that the traffic volume has not decreased or peaked. In the sixth hour, the traffic volume is `f'. Thus, it belongs to the above-medium level. Moreover, the traffic volume can be ¬a, ¬b, and Null in the next four hours, indicating that the traffic volume has not decrease significantly in these four hours and remains in the medium or above-medium levels. In the 11$^{\mathrm{th}}$ hour, the traffic volume is `c', so it belongs to the medium level. These patterns indicate that when the initial traffic volume is in the above-medium level, the traffic volume will continue to be medium or above-medium for nearly ten hours.

Therefore, we can use these negative patterns to make a relatively reasonable prediction of future traffic trends. This case study shows that negative patterns can reveal more interesting information than positive patterns in some cases.

\section{Conclusion}
The classical negative SPM methods ignore the repetition of patterns and gap constraints. Therefore, it is difficult to discover negative patterns in datasets with long sequences. To tackle this problem, this paper proposed the ONP-Miner algorithm, which has two key parts, support calculation and candidate pattern generation. To calculate the supports of negative sequential patterns, we propose the MatchDB algorithm, which employs the depth-first and backtracking strategies and can effectively avoid creating redundant nodes and parent-child relationships. To generate candidate patterns, we employ the pattern join and pruning strategies to generate and further prune the candidate patterns, respectively. The experimental results show that ONP-Miner not only is more effective, but also has better mining performance than the state-of-the-art algorithms. More importantly, we use ONP-Miner to mine ONPs in traffic volume data, and the results show that ONP mining can find more interesting patterns that can be further used  to predict future traffic.

In this paper, we investigate the one-off negative SPM which can find frequent ONPs with gap constraints. However, this research has some limitations shown as follows.

\begin{enumerate}
\item We know that users may pay more attentions to the high utility patterns \cite{  Wu 1nonHANP2021, Tin 2high2021}, rather than frequent patterns. Therefore, inspired by high utility pattern mining, high utility ONP mining is worth considering. Moreover,  three-way pattern mining methods \cite{Wu 2tri2021, wu2022tmis} were proposed, which categorize the events into three levels of interest: strong, medium, and weak interest, and patterns are composed of strong and medium interest events. Thus, the mined patterns are more concerned by users. Inspired by three-way pattern mining, three-way ONP mining is worth exploring.

\item Gap constraint is suitable for users with prior knowledge. However, in many tasks such as sequence classification \cite{Wu 4con2021}, it is difficult for users to give a rational gap constraint without prior knowledge. Therefore, ONP mining with self-adaptive gap is worth investigating.

\item Compared with other state-of-the-art algorithms, such as e-NSP \cite{Cao  6neg2016ensp} and e-RNSP \cite{Dong  7neg2020ernsp}, ONP-Miner algorithm is less efficient in mining negative patterns when mining similar number of frequent positive patterns. Therefore, further research is needed to improve the efficiency of the algorithm.
\end{enumerate}

\section*{Acknowledgement}
This work was partly supported by National Natural Science Foundation of China (61976240, 52077056, 91746209, 62120106008),  National Key Research and Development Program of China (2016YFB1000901), and Natural Science Foundation of Hebei Province, China (Nos. F2020202013, E2020202033).


\begin{thebibliography}{1}

\bibitem{gan 1survey2019}
Wensheng Gan, Jerry Chun-Wei Lin, Philippe Fournier-Viger, Han-Chieh Chao, and Philip S. Yu. 2019. A survey of parallel sequential pattern mining. ACM Transactions on Knowledge Discovery from Data 13, 3, 1-34.

\bibitem{Phi 2survey2017}
Philippe Fournier-Viger, Jerry Chun-Wei Lin, Rage Uday Kiran, Yun Sing Koh, and Rincy Thomas. 2017. A survey of sequential pattern mining. Data Science and Pattern Recognition 1, 1, 54-77.

\bibitem{wuxindong2022}
Xindong Wu, Xingquan Zhu, and Minghui Wu. 2022.The evolution of search: Three computing paradigms. ACM Transactions on Management Information Systems, 13, 2, 20, doi:10.1145/3495214.



\bibitem{pengfeizhang}
Pengfei Zhang, Tianrui Li, Zhong Yuan, Chuan Luo, Guoqiang Wang, Jia Liu, and Shengdong Du. 2022. A data-level fusion model for unsupervised attribute selection in multi-source homogeneous data. Information Fusion, 80, 87-103.



\bibitem{wanglizhen2022}
Lizhen Wang, Yuan Fang, and Lihua Zhou. 2022. Preference-based spatial co-location pattern mining, Series Title: Big Data Management. Springer Singapore, 2022, https://doi.org/10.1007/978-981-16-7566-9



\bibitem{Wu 2nonNOSEP2018}
Youxi Wu, Yao Tong, Xingquan Zhu, and Xindong Wu. 2018. NOSEP: Nonoverlapping sequence pattern mining with gap constraints. IEEE Transactions on Cybernetics 48, 10, 2809-2822.

\bibitem{Wu 1nonHANP2021}
Youxi Wu, Meng Geng, Yan Li, Lei Guo, Zhao Li, Philippe Fournier-Viger, Xingquan Zhu, and Xindong Wu. 2021. HANP-Miner: High average utility nonoverlapping sequential pattern mining. Knowledge-Based Systems 229, 107361.



\bibitem{Tin 2high2021}
Tin Truong, Hai Duong, Bac Le, Philippe Fournier-Viger, Unil Yun, and Hamido Fujita. 2021. Efficient algorithms for mining frequent high utility sequences with constraints. Information Sciences 568, 239-264.


\bibitem{weisong2021}
Wei Song, Lu Liu, and Chaomin Huang. 2021. Generalized maximal utility for mining high average-utility itemsets.  Knowledge and Information Systems, 63, 2947–2967.


\bibitem{wenshenggan2021}
Wensheng Gan, Jerry Chun-Wei Lin, Philippe Fournier-Viger, Han-Chieh Chao, and Philip S. Yu. 2021.  Beyond frequency: Utility mining with varied item-specific minimum utility. ACM Transactions on Internet Technology, 21, 1, 3.


\bibitem{Wang2018colocation}
Lizhen Wang, Xuguang Bao, Lihua Zhou. 2018. Redundancy reduction for prevalent co-location patterns. IEEE Transactions on Knowledge and Data Engineering, 30,  1,  142-155.


\bibitem{Wu 4con2021}
Youxi Wu, Yuehua Wang, Yan Li, Xingquan Zhu, and Xindong Wu. 2021. Top-k self-adaptive contrast sequential pattern mining. IEEE Transactions on Cybernetics, doi:10.1109/TCYB.2021.3082114. 

\bibitem{Rong 1con2019}
Ronghui Wu, Qing Li, and Xiangtao Chen. 2019. Mining contrast sequential pattern based on subsequence time distribution variation with discreteness constraints. Applied Intelligence 49, 12, 4348-4360.

\bibitem{Qing 2con2020}
Qingzhe Li, Liang Zhao, Yi-Ching Lee, and Jessica Lin. 2020. Contrast pattern mining in paired multivariate time series of a controlled driving behavior experiment. ACM Transactions on Spatial Algorithms and Systems 6, 4, 1-28.


\bibitem{David 3con2017}
David Savage, Xiuzhen Zhang, Pauline Chou, Xinghuo Yu, and Qingmai Wang. 2017. Distributed mining of contrast patterns. IEEE Transactions on Parallel and Distributed Systems 28, 7, 1881-1890.



\bibitem{Hu OPP2022}
Youxi Wu, Qian Hu, Yan Li, Lei Guo, Xingquan Zhu, and Xindong Wu. 2022. OPP-Miner: Order-preserving sequential pattern mining for time series. IEEE Transactions on Cybernetics, doi: 10.1109/TCYB.2022.3169327.

\bibitem{Jerry 1close2021}
Jerry Chun-Wei Lin, Youcef Djenouri, and Gautam Srivastava. 2021. Efficient closed high-utility pattern fusion model in large-scale databases. Information Fusion 76, 122-132.

\bibitem{Bac 2close2017}
Bac Le, Hai Duong, Tin Truong, and Philippe Fournier-Viger. 2017. FGenSM: Two efficient algorithms for mining frequent closed and generator sequences using the local pruning strategy. Knowledge and Information Systems 53, 71–107.




\bibitem{Wu 3close2020}
Youxi Wu, Changrui Zhu, Yan Li, Lei Guo, and Xindong Wu. 2020. NetNCSP: Nonoverlapping closed sequential pattern mining. Knowledge-Based Systems 196, 105812.

\bibitem{Zeng 1pos2019}
Zengyou He, Simeng Zhang, and Jun Wu. 2019. Significance-based discriminative sequential pattern mining. Expert Systems with Applications 122, 54-64.

\bibitem{Ting 2pos2020}
Tingting Wang, Lei Duan, Guozhu Dong, and Zhifeng Bao. 2020. Efficient mining of outlying sequence patterns for analyzing outlierness of sequence data. ACM Transactions on Knowledge Discovery from Data 14, 5, 62, 1-26.

\bibitem{Cao 1neg2021}
Wei Wang and Longbing Cao. 2021. VM-NSP: Vertical negative sequential pattern mining with loose negative element constraints. ACM Transactions on Information Systems 39, 2, 1-27.

\bibitem{Cao 2neg2019}
Wei Wang and Longbing Cao. 2019. Negative sequence analysis: a review. ACM Computing Surveys 52, 2, 32.

\bibitem{Xu campus2018}
Tiantian Xu, Tongxuan Li, and Xiangjun Dong. 2018. Efficient high utility negative sequential patterns mining in smart campus. IEEE Access 6, 23839-23847.


\bibitem{Xu msNSP2017}
Tiantian Xu, Xiangjun Dong, Jianliang Xu, and Yongshun Gong. 2017. E-msNSP: Efficient negative sequential patterns mining based on multiple minimum supports. International Journal of Pattern Recognition and Artificial Intelligence 31, 2, 1-17.

\bibitem{Jen progre2020}
Jen-Wei Huang, Yongbin Wu, and Bijay Prasad Jaysawal. 2020. On mining progressive positive and negative sequential patterns simultaneously. Journal of Information Science and Engineering 36, 1, 145-169.

\bibitem{pmdb2022}
Philippe Fournier-Viger, Wensheng Gan, Youxi Wu, Mourad Nouioua, Wei Song, Tin Truong, and Hai Duong. 2022. Pattern mining: Current challenges and opportunities.  1st Workshop on Pattern mining and Machine learning in Big complex Databases, 1-16.

\bibitem{Chunkai 4top-k2021}
Chunkai Zhang, Zilin Du, Wensheng Gan, and Philip S. Yu. 2021. TKUS: Mining top-k high utility sequential patterns. Information Sciences 570, 342-359.

\bibitem{Jen 5top-k2019}
Jen-Wei Huang, Bijay Prasad Jaysawal, Kuanying Chen, and Yongbin Wu. 2019. Mining frequent and top-k high utility time interval-based events with duration patterns. Knowledge and Information Systems 61, 3, 1331-1359.

\bibitem{Fabio 4close2016}
Fabio Fumarola, Pasqua Fabiana Lanotte, Michelangelo Ceci, and Donato Malerba. 2016. CloFAST: Closed sequential pattern mining using sparse and vertical id-lists. Knowledge and Information Systems 48, 2, 429-463.

\bibitem{Md 1max2018}
Md. Rezaul Karim, Michael Cochez, Oya Deniz Beyan, Chowdhury Farhan Ahmed, and Stefan Decker. 2018. Mining maximal frequent patterns in transactional databases and dynamic data streams: A spark-based approach. Information Sciences 432, 278-300.

\bibitem{Yan 2max2021}
Yan Li, Shuai Zhang, Lei Guo, Jing Liu, Youxi Wu, and Xindong Wu. 2022. NetNMSP: Nonoverlapping maximal sequential pattern mining. Applied Intelligence 52, 9, 9861-9884.

\bibitem{Fan 1tri2020}
Fan Min, Zhiheng Zhang, Wenjie Zhai, and Rongping Shen. 2020. Frequent pattern discovery with tri-partition alphabets. Information Sciences 507, 715-732.

\bibitem{Wu 2tri2021}
Youxi Wu, Lanfang Luo, Yan Li, Lei Guo, Philippe Fournier-Viger, Xingquan Zhu, and Xindong Wu. 2022. NTP-Miner: Nonoverlapping three-way sequential pattern mining. ACM Transactions on Knowledge Discovery from Data 16, 3, 51.


\bibitem{wu2022tmis}
Youxi Wu, Xiaohui Wang, Yan Li, Lei Guo, Zhao Li, Ji Zhang, and Xindong Wu. 2022. OWSP-Miner: Self-adaptive one-off weak-gap strong pattern mining. ACM Transactions on Management Information Systems 13, 3, 25.

\bibitem{wu2022ins}
Youxi Wu, Zhu Yuan, Yan Li, Lei Guo, Philippe Fournier-Viger, and Xindong Wu. 2022. NWP-Miner: Nonoverlapping weak-gap sequential pattern mining. Information Sciences 588, 124-141. 


\bibitem{Yoo 3high2021}
Yoonji Baek, Unil Yun, Heonho Kim, Jongseong Kim, Bay Vo, Tin C. Truong, and Zhihong Deng. 2021. Approximate high utility itemset mining in noisy environments. Knowledge-Based Systems 212, 106596.

\bibitem{Yoo 4high2021}
Yoonji Baek, Unil Yun, Heonho Kim, Hyoju Nam, Hyunsoo Kim, Jerry Chun-Wei Lin, Bay Vo, and Witold Pedrycz. 2021. RHUPS: Mining recent high utility patterns with sliding window-based arrival time control over data streams. ACM Transactions on Intelligent Systems and Technology 12, 2, 1-27.




\bibitem{wu2021eswa}
Youxi Wu, Rong Lei, Yan Li, Lei Guo, Xindong Wu. 2021. HAOP-Miner: Self-adaptive high-average utility one-off sequential pattern mining. Expert Systems With Applications 184, 115449.


\bibitem{Unil dyn2020}
Unil Yun, Hyoju Nam, Jongseong Kim, Heonho Kim, Yoonji Baek, Judae Lee, Eunchul Yoon, Tin C. Truong, Bay Vo, and Witold Pedrycz. 2020. Efficient transaction deleting approach of pre-large based high utility pattern mining in dynamic databases. Future Generation Computer Systems 103, 58-78.

\bibitem{Jin assoc2020}
Jinane Harmouche and Sriram Narasimhan. 2020. Long-term monitoring for leaks in water distribution networks using association rules mining. IEEE Transactions on Industrial Informatics 16, 1, 258-266.

\bibitem{Sum  inter2014}
Sumalatha Saleti and R. B. V. Subramanyam. 2020. Distributed mining of high utility time interval sequential patterns using mapreduce approach. Expert Systems with Applications 141, 112967.



\bibitem{Wu 1rep2014}
Youxi Wu, Lingling Wang, Jiadong Ren, Wei Ding, and Xindong Wu. 2014. Mining sequential patterns with periodic wildcard gaps. Applied Intelligence 41, 1, 99–116.


\bibitem{Wamg 3non2022}
Yuehua Wang, Youxi Wu, Yan Li, Fang Yao, Philippe Fournier-Viger, Xindong Wu. 2022. Self-adaptive nonoverlapping sequential pattern mining. Applied Intelligence 52(6): 6646-6661.

\bibitem{Fei  gap2014}
Fei Xie, Xindong Wu, and Xingquan Zhu. 2017. Efficient sequential pattern mining with wildcards for keyphrase extraction. Knowledge-Based Systems 115, 27–39. 




\bibitem{TCBB2020gap}
Yu-Hao Ke, Jen-Wei Huang, Wei-Chen Lin, Bijay Prasad Jaysawal. 2020. Finding possible promoter binding sites in DNA sequences by sequential patterns mining with specific numbers of gaps. IEEE/ACM Transactions on Computational Biology and Bioinformatics 18, 6, 2459-2470.


\bibitem{ganneg2022}
Gengsen Huang, Wensheng Gan, Shan Huang, Jiahui Chen. 2022. Negative pattern discovery with individual support. Knowledge-Based Systems, 251,109194.


\bibitem{Dong  1neg2019}
Xiangjun Dong, Ping Qiu, Jinhu Lv, Longbing Cao, and Tiantian Xu. 2019. Mining top-k useful negative sequential patterns via learning. IEEE Transactions on Neural Networks and Learning Systems 30, 9, 2764-2778.

\bibitem{Gao  2neg2021}
Xinming Gao, Yongshun Gong, Tiantian Xu, Jinhu Lu, Yuanhai Zhao, and Xiangjun Dong. 2021. Toward to better structure and looser constraint to mine negative sequential patterns. IEEE Transactions on Neural Networks and Learning Systems, doi: 10.1109/TNNLS.2020.3041732.

\bibitem{Ping  3neg2021}
Ping Qiu, Yongshun Gong, Yuhai Zhao, Longbing Cao, Chengqi Zhang, and Xiangjun Dong. 2021. An efficient method for modeling nonoccurring behaviors by negative sequential patterns with loose constraints. IEEE Transactions on Neural Networks and Learning Systems, doi:10.1109/TNNLS.2021.3063162.



\bibitem{Cao  6neg2016ensp}
Longbing Cao, Xiangjun Dong, and Zhigang Zheng. 2016. e-NSP: Efficient negative sequential pattern mining. Artificial Intelligence 235, 156-182.

\bibitem{Dong  7neg2020ernsp}
Xiangjun Dong, Yongshun Gong, and Longbing Cao. 2020. e-RNSP: An efficient method for mining repetition negative sequential patterns. IEEE Transactions on Cybernetics 50, 5, 2084-2096.

\bibitem{Yu  SRMP2021}
{Yan Li, Lei Yu, Jing Liu, Lei Guo, Youxi Wu, Xindong Wu. 2021. NetDPO: (delta, gamma)-approximate pattern matching with gap constraints under one-off condition. Applied Intelligence, doi: 10.1007/s10489-021-03000-2.}

\bibitem{Wu  nettree2017}
{Youxi Wu, Cong Shen, He Jiang, Xindong Wu. 2017. Strict pattern matching under non-overlapping condition. Science China Information Sciences 60 (1):012101.}

	
\end{thebibliography}
\end{document}